\documentclass{sig-alternate}
\usepackage{graphicx}
\usepackage{subfigure}
\usepackage[lined,boxed,commentsnumbered, ruled]{algorithm2e}
\usepackage{times,amsmath,epsfig,bm}
\usepackage{hyperref}
\usepackage{breakurl}
\usepackage{multirow}
\usepackage{array}
\usepackage{verbatim}

\newdef{myDef}{Definition}

\begin{document}
%

\title{Geo-SAGE: A Geographical Sparse Additive Generative Model for Spatial Item Recommendation}

\author{Weiqing Wang{\small$^\dag$} \hspace*{7pt} Hongzhi Yin{\small$^\dag$} \hspace*{7pt} Ling Chen{\small$^\ddag$}\hspace*{7pt}  Yizhou Sun{\small$^\S$} \hspace*{7pt} Shazia Sadiq{\small$^\dag$} \hspace*{7pt} Xiaofang Zhou{\small$^\dag$}\\
  \affaddr{$~^\dag$The University of Queensland, School of Information Technology and Electrical Engineering, Qld 4072, Australia}\\
\affaddr{$~^\S$College of Computer and Information Science, Northeastern University}\\
\affaddr{ $~^\ddag$QCIS, University of Technology, Sydney}\\
 \email{$~^\dag$\{weiqingwang, h.yin1, uqssadiq, uqxzhou\}@uq.edu.au,}\\ \email{$~^\S$yzsun@ccs.neu.edu,} \email{$~^\ddag$ling.chen@uts.edu.au}}

\maketitle
\begin{abstract}
With the rapid development of {location-based social networks (LBSNs)}, spatial item recommendation has become an important means to help people discover attractive and interesting venues and events, especially when users travel out of town. However, this recommendation is very challenging compared to the traditional recommender systems. A user can visit only a limited number of spatial items, leading to a very sparse user-item matrix. Most of the items visited by a user are located within a short distance from where he/she lives, which makes it hard to recommend items when the user travels to a far away place. Moreover, user interests and behavior patterns may vary dramatically across different geographical regions. In light of this,  we propose Geo-SAGE, a geographical sparse additive generative model for spatial item recommendation in this paper. Geo-SAGE considers both user personal interests and the preference of the crowd in the target region,  by exploiting both the co-occurrence pattern of spatial items and the content of spatial items. To further alleviate the data sparsity issue, Geo-SAGE exploits the geographical correlation by smoothing the crowd's preferences over a well-designed spatial index structure called spatial pyramid. We conduct extensive experiments to evaluate the performance of our Geo-SAGE model on two real large-scale datasets. The experimental results clearly demonstrate our Geo-SAGE model outperforms the state-of-the-art in the two tasks of both out-of-town  and home-town  recommendations.

\end{abstract}
\vspace{-8pt}
\section{introduction}
The rapid development of Web 2.0, location acquisition and wireless
communication technologies have fostered a number of
\emph{location-based social networks (LBSNs)}, such as Foursquare,
Gowalla, Facebook Places and Loopt, where users can check-ins at venues and  share life
experiences in the physical world  via
mobile devices~\cite{bao:location}. On one hand, the new dimension
of location implies extensive knowledge about an individual's
behavior and interests by bridging the gap between online social
networks and the physical world, which enables us to better
understand user preferences and design optimal recommendation
systems. On the other hand, it is valuable to develop the
\textit{location recommendation service} as an essential function to
LBSNs to encourage users to explore new locations~\cite{Ference:2013:LRO:2505515.2505637}.
Therefore, developing recommendation systems for LBSNs to provide
users with spatial items (e.g., a venue or an event associated with
a geographic location) has recently attracted increasing research
attention~\cite{bao2014survey,Lian:2014:GJG:2623330.2623638}.
This application becomes more important and useful when a user
travels to an unfamiliar area, where he/she has little knowledge
about the neighbourhood. In this scenario, the recommender
system is proposed as  \emph{recommendation for out-of-town users in
LBSNs} in \cite{Ference:2013:LRO:2505515.2505637}. In this paper, we focus on the
problem of spatial item recommendation, aiming to offer accurate
recommendations for both home-town and out-of-town users by mining
their historical behavior data in LBSNs.

Spatial item recommendation 
is a highly challenging problem because of the following three main
reasons. 1. \textit{Data Sparsity}. While LBSNs expand rapidly, the
number of spatial items visited by an individual user is rather small
compared to the total number of spatial items in a LBSN, which results in
a very sparse user-spatial item matrix. This issue plagues most of the existing
collaborative filtering recommender systems. 2. \textit{Travel Locality}.
As observed in \cite{levandoski:lars}, users tend to visit spatial
items that are nearby to their home
locations. 
For example, an analysis of Foursquare data shows that 45\% of users
travel 10 miles or less, while 75\% travel 50 miles or less
\cite{levandoski:lars}. Travel locality makes the out-of-town
recommendation more challenging. On one hand, there may be few or no
historical activity  records of the target user around the current
location. On the other hand,  similar users to the target user (that is,
those who exhibit similar spatial item visiting behaviors) may also have not visited spatial items near
the current location. Again, collaborative filtering based
approaches will be incapable of providing effective recommendations.
3. \textit{Interest Drift Across Regions}. When a user travels to a different
region, her interests or behavior patterns may change. Through
Foursquare API\footnote{\url{https://developer.foursquare.com/}}, we
extract the top three categories of check-in POIs (points of interest) in three different cities including Boston, Las Vegas and the Gold
Coast for a group of users. The group size is 3000 and each user of this group has
check-in records in all three cities. The percentage of check-ins of
each POI category is shown in Table ~\ref{tbl:interestdrift}. We
observe that when users are in Las Vegas, they are more interested
in visiting Casino (80.32\%), Nightlife (10.61\%) and Outlet
(5.82\%), while the same users prefer Beach (71.36\%), Surf Spot(14.82\%) and Theme Park (9.60\%) when in Gold Coast. A good spatial
item recommender system should effectively model user interest drift
and take it into account when making recommendations for out-of-town
users.

Recently, several methods
\cite{bao:location,Ference:2013:LRO:2505515.2505637,Yin:2013:LLR:2487575.2487608} have been developed
to make recommendations for both home-town and out-of-town users. As a matter of fact, these approaches either do not address
all three challenges for spatial item recommendation, or address the challenges with ineffective strategies. For example,
Ference et al.~\cite{Ference:2013:LRO:2505515.2505637} proposed a CF-based method
that considers both users who have visited many common spatial items
with the target user and friends mined from the target user's social
connection in LBSNs. Including visiting records of social friends is supposed to handle data sparsity, and also the travel locality.
For example, when similar users cannot provide effective clues for
out-of-town recommendations, the visiting records of social friends
will be used. Nonetheless, according to the survey \cite{cho:friendship}, when the user travels more than 100km, the check-in probability at the same place visited by any of the social ties, is merely 10\%. Bao
et al.~\cite{bao:location} presented a recommender system considering
both user personal interests and opinions from local experts. They
model individual user's interests based on the category information
of spatial items, which helps overcome \textit{data sparsity} to
some extend. The opinions of local experts are used to solve the
\emph{travel locality} problem. Unfortunately, since the chosen local
experts are users who share same/similar interests with the target user,
this approach cannot address \emph{interest drift}.

\begin{table}[t]
\small \centering
\begin{tabular}{|c|c|c|}
      \hline
      City & Top POI Types & Percentage of Check-ins(\%) \\\hline
       \multirow{3}{*}{Gold Coast (AU)} & Beach & 71.36\% \\\cline{2-3}
       & Surf Spot & 14.82\% \\\cline{2-3}
       & Theme Park  & 9.60\% \\\cline{2-3}\hline
      \multirow{3}{*}{Las Vegas (US)} & Casino & 80.32\% \\\cline{2-3}
       & Nightlife & 10.61\% \\\cline{2-3}
       & Outlet & 5.82\% \\\cline{2-3}\hline
       \multirow{3}{*}{Boston (US)} &  College  & 78.32\% \\\cline{2-3}
       &Museum  & 9.45\% \\\cline{2-3}
        &Park  & 7.65\% \\\cline{2-3}
      \hline
    \end{tabular}
    \vspace{-4pt}
\caption{Illustration of User Interest Drift}
\label{tbl:interestdrift}
\vspace{-17pt}
\end{table}

In this paper, we propose Geo-SAGE, a geographical sparse additive
generative model for spatial item recommendation. Traditional
mixture models, such as~\cite{Yin:2013:LLR:2487575.2487608}, consider multiple facets
(e.g., time and location) that influence user choosing of spatial
items by introducing additional latent variables, acting as
``switches'', to control which facet is currently active. As a matter of fact, it
is not only computationally expensive to learn personalized
``switching'' variables for individual users but also difficult to
learn these variables accurately given sparse datasets. Inspired by
the the Sparse Additive Generative model
(SAGE)~\cite{eisenstein:sparse}, we have designed our similar model by adding the effect of all the facets in
the exponential space to avoid the inference of the latent
``switching'' variables, aiming to achieve improved robustness and
predictive accuracy.

Basically, to model user visiting behaviors, Geo-SAGE takes into
account both user interests and interest drift. Geo-SAGE learns user
interests as distributions over a set of latent topics, by mining
both the co-occurrence patterns of spatial items  and their content
information (e.g., tags and categories). Exploiting
content information of spatial items not only alleviates the data
sparsity issue but also addresses the travel locality for out-of-town
recommendation. The content of spatial items serves as the medium to
transfer user interests learned from home town to  unfamiliar regions.

To adapt to \emph{interest drift} across regions, Geo-SAGE recognizes two roles of an
individual user in a spatial region: a local or a tourist. Given a location, visiting records from local
users will be mined to learn \textit{native preference} as a
distribution over latent topics. Similarly, visiting records from
tourists will be used to learn \textit{tourist preference}. Users with the same role at a location are more likely to have similar preferences and behavior patterns.  Thus, to recommend spatial items to a target user $u$ at location $l$, we  consider not only $u$'s personal interests, but also the preference and visiting behaviors of the crowd who have the same role with $u$ at location $l$.

However, by using the visiting history to model native
preference and tourist preference, we face the data sparsity issue
again, especially when the target region is small. When there are not
enough visiting records in the target region, the two variables cannot be
inferred accurately. To further overcome the data sparsity, Geo-SAGE  integrates the
spatial index - \emph{spatial pyramid} which is a tree structure
proposed in \cite{levandoski:lars}. It is first constructed by
partitioning locations of spatial items into spatial grids of
varying sizes at different hierarchies. Then, Geo-SAGE applies the \emph{additive}
framework \cite{ahmed:latent,kanagal:supercharging} to learn native
preference and tourist preference of each region. Briefly, when
learning the native preference and tourist preference in a region,
the two variables learned for all of the region's ancestor grids in
the spatial pyramid will be added.   By doing this, if there are few or no activities
in a region, we can still infer its native preference and tourist preference guided by its ancestors. By integrating with the spatial
pyramid, Geo-SAGE gains another advantage by allowing users to switch
between different scales of geo-regions (e.g., zoom in/out on a Google
Map) without re-learning the parameters. In this way, this model can
be connected with Google Map seamlessly.

The main contributions of our work are summarized as follows.
\begin{itemize}
	\vspace{-3pt}
	\item  We design a novel geographical sparse additive generative model (Geo-SAGE) for spatial item recommendation which incorporates and exploits item content information and  the crowd's preference at a region to address the problems of data sparsity, travel locality and interest drift across regions.
	\vspace{-3pt}
	\item We refine the crowd's preference at a region by distinguishing native preference from tourist preference,  based on which  Geo-SAGE exploits the geographical correlation by smoothing these preferences over a well-designed spatial index structure - spatial pyramid, to further alleviate the data sparsity .
	\item We conduct extensive experiments to evaluate the performance of the proposed Geo-SAGE  on two real large-scale datasets. The experimental results clearly demonstrate that Geo-SAGE outperforms the state-of-the-art in the tasks of both out-of-town and home-town recommendations
	\vspace{-4pt}
\end{itemize}

This paper is organized as follows: Section 2 provides preliminary
concepts about the SAGE model. Section 3 formulates our spatial
item recommendation problem and describes the Geo-SAGE model.
Section 4 discusses the experimental evaluation of Geo-SAGE.
Existing research related to our work is surveyed in Section 5.
Section 6 closes this paper with some conclusive remarks.

\vspace{-8pt}
\section{Preliminaries about SAGE}
Our model is inspired by the Sparse Additive Generative
Model\\(SAGE) \cite{eisenstein:sparse}, which has been shown to be effective
generative models without explicit switching variables
\cite{hong:discovering,Hu:2013:STM:2507157.2507174,hu:spatio-temporal}.  The basic
idea of the  model is that, if a variable is affected by several
components, it can be generated by the mixture of these components
without any explicit indicator variable. The key difference to
traditional mixture models is that the mixture occurs in terms of natural parameters of the exponential family instead of distributions. Such a model is robust given limited training data
as it does not have to infer a complex indicator variable
distinguishing the set of causes.

To provide a clearer explanation of SAGE, we use a traditional probabilistic mixture generative model, LCA-LDA~\cite{Yin:2013:LLR:2487575.2487608}, as an illustrative example. LCA-LDA, is a location-content-aware model that aims to mimic the
process of human decision making on spatial items. The
model considers the user's personal interest $\theta^{user}_{u}$ and
the influence of local preference $\theta^{crowd}_{l}$ (note that it does not distinguish between native preference and tourist preference) in a unified
manner, and automatically leverages the effect of the two factors.
Specifically, given a querying user $u$ at a target location $l$, the
likelihood that user $u$ will prefer item $v$ is computed by combining these two factors through a linear
combination of them as follows.
\begin{equation}
\small \label{eq:ulalda} P(v|\theta^{user}_{u},\theta^{crowd}_{l})
= \lambda_u P(v|\theta^{user}_u) +
(1-\lambda_{u})P(v|\theta^{crowd}_{l})
\end{equation}
where $\lambda_u$ is the ``switching'' variable that needs to be inferred for each user. Obviously, it cannot be inferred accurately when the training data for the individual user is sparse.
In contrast, SAGE combines the two generative facets through simple addition in exponential space as shown in Equation~\ref{equ:SAGE}. Clearly, it avoids the need for  latent switching variables.
\begin{equation}
\vspace{-2pt} \small \label{equ:SAGE}
\begin{split}
P(v|\theta^{user}_{u},\theta^{crowd}_{l})& = P(v|\theta^{user}_{u}+\theta^{crowd}_{l})\\
&= \frac{exp(\theta^{user}_{u,v}+\theta^{crowd}_{l,v})}{\sum_{v'}
exp(\theta^{user}_{u,v'}+\theta^{crowd}_{l,v'})}
\end{split}
\vspace{-2pt}
\end{equation}

\vspace{-8pt}
\section{Geographical SAGE Model}

In this section, we first formulate the problem definition, and then
present our proposed geographical SAGE model (Geo-SAGE).

\subsection{Problem Definitions}

For ease of  presentation, we define the key data structures and
notations used in this paper. Table \ref{tab:input} also lists them.

\newdef{definition3}{Definition}
\begin{definition3}
\vspace{-4pt}
 \textbf{(Spatial Item)}
 A spatial item is an item associated with a geographical location (e.g., a restaurant or a cinema).
 \vspace{-5pt}
\end{definition3}

 In our model, a spatial item has three attributes: identifier,  location and contents.
 We use $v$ to represent a spatial item identifier, $l_v$ to denote its corresponding location identifier and $W_v$ to represent the set of words describing the semantic information of the item (e.g., tags and categories). POI is a kind of spatial item. The location information available for each spatial item $v$ in the collected raw datasets is in the form of the (latitude, longitude) pair. Then a spatial pyramid structure~\cite{levandoski:lars,sarwati:lars}, is applied to partition and index  the entire geographic area. The granularities depend on the nature of the application and range from cities to streets.  The details of the spatial pyramid are described in subsection $3.4$.

\begin{definition3}
 \vspace{-4pt}
 \textbf{(User Home Location)}
   Following the recent work of ~\cite{Li:2012:TSU:2339530.2339692}, given a user $u$, we define the user's home location as the place where the user lives, denoted as $l_u$.
 \vspace{-5pt}
\end{definition3}

Note that, we assume a user's home location is ``permanent'' in our problem. In other words, a home location is a static location instead of a real-time location  that is ``temporally'' related to him (e.g., the places where she/he is traveling). Due to privacy, user home locations are not always available. For
a user whose home location is not explicitly given, we adopt the method similar to~\cite{conf/icwsm/ScellatoNLM11} by inferring the user's home location as the cell in the spatial pyramid with the most of his/her check-ins.

\begin{definition3}
\vspace{-4pt}
 \textbf{(User Activity)}
 A user activity  is made of a  five tuple $(u,$ $~v,$ $~l_v,$ $~W_v$, $s)$ which indicates that the user $u$ visits the spatial item $v$, located at $l_v$ and described as $W_v$, in the role of $s$.
  If $s=0$, the user is recognized as a local and the activity occurs in $u$'s home town. If $s=1$, the user $u$ plays the role of tourist when visiting $v$.
 \vspace{-5pt}
\end{definition3}
\begin{definition3} \textbf{(User Profile)}
For each user $u$,  we create a user profile $D_u$, which is a set of user  activities associated with $u$.  The dataset $D$ used in our model consists of user profiles, that is, $D=\{D_u:u\in U\}$.
  \vspace{-5pt}
\end{definition3}

Then, given a dataset $D$ as the union of a collection of user profiles, we aim to provide spatial item recommendation for both home-town and out-of-town users. We formulate our problem that takes into account both of the two scenarios in a unified fashion as follows.
\vspace{-2pt}
\newtheorem{problem}{Problem}
\begin{problem} \textbf{(Spatial Item Recommendation)}
 Given a user activity dataset $D$ and  a target user $u$ with his/her current location $l$ (that is, the query is $q=(u,~l)$), our goal is to
 recommend a list of spatial items that $u$ would be interested in. Given a distance threshold $d$, the problem becomes an \textbf{out-of-town recommendation} if the distance between the target user's current location and his/her home location (that is, $|l-l_u|$) is greater than $d$. Otherwise, the problem is a  \textbf{home-town recommendation}.
  \vspace{-3pt}
\end{problem}

Following related studies~\cite{Ference:2013:LRO:2505515.2505637,mok2010does}, we set
$d=100km$ in our work, since a distance around $100km$ is the typical radius of human ``reach'' $-$ it takes about 1 to 2 hours to drive such a distance.

\begin{table}[!t]
\small
\centering
\begin{tabular}{>{\centering} p{35pt} |>{\centering} p{180pt}} \hline
Variable & Interpretation \tabularnewline \hline $U,R,V$ & the set
of users, locations and spatial items\tabularnewline \hline $W$ &
the vocabulary set \tabularnewline \hline $D_u$ & the
profile of user $u$ \tabularnewline \hline $v_{u,i}$ & the spatial
item of $i^{th}$ record in $D_u$ \tabularnewline \hline $l_{u,i}$ &
the location of spatial item $v_{u,i}$ \tabularnewline \hline $l_u$
& the home location of the user $u$ \tabularnewline
\hline $W_{u,i}$ & the set of  words describing spatial item
$v_{u,i}$ \tabularnewline \hline $w_{u,i,n}$ & the $n^{th}$ content
word describing spatial item $v_{u,i}$ \tabularnewline \hline
$s_{u,i}$ & if the user $u$ is a local or tourist on location
$l_{u,i}$\tabularnewline \hline
\end{tabular}
\caption{Notations of The Input Data}
\label{tab:input}
\vspace{-6mm}
\end{table}

\subsection{Model Description}

\begin{figure}[!t]
   \centering
    \small
   \includegraphics[width=0.6\columnwidth]{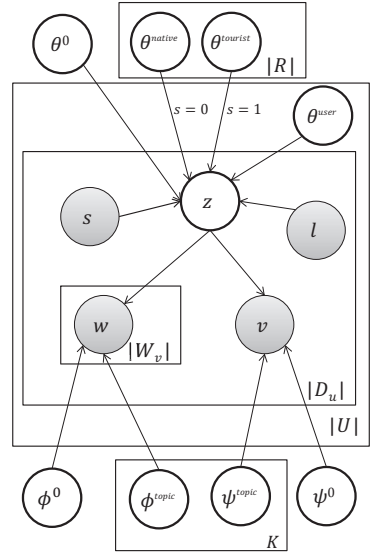}
\vspace{-4mm}
\caption{The Graphical Representation of Our Model}
\label{fig:model}
\vspace{-4mm}
\end{figure}

To model user activities, we propose a geographical sparse additive generative model (Geo-SAGE).  Figure~\ref{fig:model} shows the
graphical representation of Geo-SAGE.  We first introduce the notations of our model, which are listed in Table~\ref{tab:parameters}.
Our input data, that is, users' activity profiles, are modeled as observed random variables, shown as shaded circles in Figure~\ref{fig:model}.
Similar to existing models~\cite{Hu:2013:STM:2507157.2507174,Yin:2013:LLR:2487575.2487608}, the topic index of each user activity  is considered as a latent random variable, which is denoted as $z$.

Intuitively, a user chooses a spatial item at a given location by matching his/her personal interests with the content of that item.  Inspired by the early
work on user interest modeling~\cite{Hu:2013:STM:2507157.2507174,Liu:2013:SDM,Yin:2013:LLR:2487575.2487608},
Geo-SAGE also adopts latent topics  to characterize  users' interests.
Specifically, we infer an individual user's interest distribution over
a set of topics according to his/her visited spatial items and their associated contents, denoted as
$\theta^{user}_{u}$.  Thus, our model alleviates the data sparsity,
especially for the out-of-town recommendation, as the content of spatial items plays the role of medium through which user interests inferred
from their home town can be transferred to out-of-town regions.
Besides, we also introduce a background distribution over topics
$\theta^{0}$ to capture common topics among all users. The purpose
of using a background model $\theta^{0}$ is to make the user
interests $\theta^{user}_{u}$  learned from the dataset more
discriminative.

To adapt to user interest drift across regions, we exploit the preference of the crowds who share the same role with the target user $u$. For example, the preference of the tourists will be leveraged if the target user is currently out-of-town.  Technically, we introduce two parameters: \emph{native preference} and \emph{tourist preference}.  Given a location $l$, the native preference represents the preference of people living at location $l$, denoted as $\theta^{native}_l$. In contrast, the tourist preference represents the preference of tourists travelling in location $l$, denoted as $\theta^{tourist}_l$. Both $\theta^{native}_l$ and $\theta^{tourist}_l$ are represented by distributions over topics.  Note that, distinguishing native preference from tourist preference is one of the fundamental differences between our model and the LCA-LDA model~\cite{Yin:2013:LLR:2487575.2487608} which also exploits the local activity records at the target location.

To take full advantage of the strengths of both content-based and collaborative filtering-based recommendation methods, a topic $z$ in
our Geo-SAGE model is not only associated with a word distribution
$\phi^{topic}_z$, but also a distribution over spatial items
$\psi^{topic}_z$. This design enables $\phi^{topic}_z$ and
$\psi^{topic}_z$ to be mutually influenced and enhanced during the
topic discovery process by associating them. Thus, the discovered
topic $z$, on one hand, can cluster the content-similar items
together. On the other hand, it can also capture the item co-occurrence
patterns to link relevant items together, similar to item-based
collaborative filtering methods. Besides, we also introduce two
background models for words and items, respectively: $\phi^{0}$ and
$\psi^{0}$.  The purpose of using  background models is to make the
topics learned from the dataset more discriminative, since
$\phi^{0}$ and $\psi^{0}$ assign high probabilities to
non-discriminative and non-informative words and items.

\begin{table}[!t]
\centering
\small
\begin{tabular}{>{\centering} p{35pt} |>{\centering} p{180pt}} \hline
Variable & Interpretation \tabularnewline \hline $K$ & the number of
topics \tabularnewline \hline
$z_{u,i}$ & the topic assigned to spatial item $v_{u,i}$
\tabularnewline \hline $\theta^0$ & the topic distribution of the
background \tabularnewline \hline $\theta^{user}_{u}$ & the topic
distribution, representing the intrinsic interest of user
$u$\tabularnewline \hline $\theta^{native}_{l}$ & the topic
distribution of $l$ , representing the native preference at
$l$\tabularnewline \hline $\theta^{tourist}_{l}$ & the topic
distribution of $l$, representing the tourist preference at $l$
\tabularnewline \hline $\phi^0$ & the word distribution of the
background \tabularnewline \hline $\psi^0$ & the spatial item
distribution of the background\tabularnewline \hline
$\phi^{topic}_{z}$ & content word distribution of topic $z$
\tabularnewline \hline $\psi^{topic}_{z}$ & spatial item
distribution of topic $z$ \tabularnewline \hline
\end{tabular}
\caption{Notations of Model Parameters}
\label{tab:parameters}
\vspace{-12pt}
\end{table}

In summary, our Geo-SAGE model aims to mimic the process of human
decision making on visiting spatial items.  Geo-SAGE assumes that users'
activities are influenced by three factors: $1)$ the user's intrinsic
interests $\theta^{user}_{u}$, $2)$ the tourists' preference $\theta^{tourist}$ if the user is a tourist; or the
natives' preference  $\theta^{native}$ if the user is a local, $3)$ the general public's preference
$\theta^0$.  All these factors are represented by a distribution over a set of topics, and
 influence the user's decision-making through generating a topic $z$. However, different from traditional mixture models~\cite{Yin:2013:LLR:2487575.2487608} which have to consider the multiple factors by introducing additional ``switching'' variables, our Geo-SAGE model adds the influence of the three factors in the exponential space, which is more efficient and robust by avoiding inferring ``switching'' variables.

The generative process of the Geo-SAGE model for
an individual user activity in the user profile $D_u$ is as follows.

\newcounter{TempEqCnt}
\setcounter{TempEqCnt}{\value{equation}} \setcounter{equation}{5}
  \begin{figure*}[!htb]
  \vspace{-4pt}
 \hrulefill
\begin{equation}
\label{eq:notations} \scriptsize
\alpha_{u,s,l,z}=\frac{exp(\theta^0_{z}+\theta^{user}_{u,z}+(1-s)\times\theta^{native}_{l,z}+s\times\theta^{tourist}_{l,z})}{\sum_{zz}exp(\theta^0_{zz}+\theta^{user}_{u,zz}+(1-s)\times\theta^{native}_{l,zz}+s\times\theta^{tourist}_{l,zz})},
\beta_{z,w}=\frac{exp(\phi^0_{w}+\phi^{topic}_{z,w})}{\sum_{ww}exp(\phi^0_{ww}+\phi^{topic}_{z,ww})},
\gamma_{z,v}=\frac{exp(\psi^0_z+\psi^{topic}_{z,v})}{\sum_{vv}exp(\psi^0_z+\psi^{topic}_{z,vv})}
\end{equation}

\begin{equation}
\label{equ:objectiveFunction} \scriptsize
\begin{split}
 P(\textbf{z},~\textbf{w},~\textbf{v}|\circleddash,~\textbf{s},~\textbf{u}) &= P(\textbf{z}|\textbf{s},~\textbf{u}, \theta^0, \theta^{user},  \theta^{native}, \theta^{tourist}) P(\textbf{w}|\textbf{z}, \phi^0, \phi^{topic}) P(\textbf{v}|\textbf{z}, \psi^0, \psi^{topic})\\
&=
\prod_{u=1}^{|U|}\prod_{i=1}^{|D_u|}\alpha_{u,s_{u,i},l_{u,i},z_{u,i}}
\prod_{u=1}^{|U|}\prod_{i=1}^{|D_u|}\prod_{n=1}^{|{W}_{v_{u,i}}|}\beta_{z_{u,i},w_{u,i,n}}
 \prod_{u=1}^{|U|}\prod_{i=1}^{|D_u|}\gamma_{z_{u,i},v_{u,i}}
\end{split}
\end{equation}
\hrulefill
\end{figure*}
\setcounter{equation}{\value{TempEqCnt}}

\begin{itemize}
\vspace{-4pt}
\item Draw a topic index $z_{u,i}$

$z_{u,i} \sim P(z_{u,i}|s_{u,i}, \theta^0, \theta^{user},
\theta^{native}, \theta^{tourist})$

\item For each content word $w_{u,i,n}$ in ${W}_{u,i}$, draw

$w_{u,i,n} \sim P(w_{u,i,n}|\phi^0, z_{u,i}, \phi^{topic})$

\item Draw a spatial item $v_{u,i}$

$v_{u,i} \sim P(v_{u,i}|\psi^0, z_{u,i}, \psi^{topic}) $
\vspace{-4pt}
\end{itemize}

For each user activity, Geo-SAGE first chooses the topic this activity is about. To generate the topic index $z$, we utilize a multinomial model as expressed in Equation \ref{equ:topicGen}.
\begin{equation}
\label{equ:topicGen} \small \centering
\begin{aligned}
& P(z_{u,i}|u, s_{u,i}, \theta^0 , \theta^{user}, \theta^{native},\theta^{tourist})\\
 = & P(z_{u,i}|\theta^0 + \theta^{user}_u  + (1-s_{u,i}) \times \theta^{native} _{l}+ s_{u,i} \times \theta^{tourist}_{l})
\end{aligned}
\end{equation}
where $\theta^0$ is the global distribution over topics. $\theta^{user}_u$ is a user-dependent distribution over topics which
represents the interests of $u$. $s$ is an observable variable which
only has two values: 0 and 1. If $s$ equals to 0, $u$ is a local at
$l$ and otherwise a tourist. $\theta^{native}_{l}$ denotes the
native preference while $\theta^{tourist}_{l}$ denotes the tourist
preference at $l$,  both of which are distributions over topics. $P(z|\theta^0 + \theta^{user}_u  +
(1-s) \times \theta^{native} _{l}+ s \times
\theta^{tourist}_{l})$, denoted as $\alpha_{u,s,l,z}$, can be computed as in
Equation~\ref{eq:notations}.
Once the topic $z$ is generated, the spatial item  $v$ and the associated content words are generated
as expressed in Equations~\ref{equ:itemGen} and \ref{equ:contentGen}, respectively.
\begin{equation}
\label{equ:itemGen} \small \centering
 P(v_{u,i}|\psi^0 , z_{u,i}, \psi^{topic})
=  P(v_{u,i}|\psi^0 + \psi^{topic}_{z_{u,i}})
\end{equation}
\begin{equation}
\label{equ:contentGen} \small P(w_{u,i,n}|\phi^0 ,z_{u,i},
\phi^{topic}) = P(w_{u,i,n}|\phi^0 + \phi^{topic}_{z_{u,i}})
\end{equation}
where $P(v|\psi^0 + \psi^{topic}_{z})=\gamma_{z,v}$ and $P(w|\phi^0
+ \phi^{topic}_{z})=\beta_{z,w}$ are computed as in
Equation~\ref{eq:notations}. Note that in order to model topics based on the background word/item distributions, for each topic,
Geo-SAGE models the difference from the
background word/item distribution in log-frequencies, instead of the frequencies
themselves.

\subsection{Model Inference}
Our goal is to learn parameters that maximize the marginal
log-likelihood of the observed random variables \textbf{w},
\textbf{v} and \textbf{s}, and the marginalization is performed with
respect to the latent random variable \textbf{z}. However, it is
difficult to be maximized directly. Therefore, we apply a mixture
between EM and a Monte Carlo sampler,  called Gibbs EM algorithm
\cite{wallach:topic}, to maximize the complete data likelihood in
Equation \ref{equ:objectiveFunction}, where $\circleddash$ is the
set of all the parameters. In the E-step, we sample latent topic
assignments by fixing all other parameters using Gibbs sampling. In the
M-step, we optimize model parameters $\circleddash$ by fixing all
topic assignments. The two steps are iterated until convergence.

More specifically, we iteratively draw latent topic \textbf{z}
for all activities in the E-step. According to the Gibbs
Sampling, when sampling $z_{u,i}$ as expressed in Equation \ref{equ:sampleFunction}, we assume all other variables are
fixed. $z_{\neg u,i}$
represents the topic assignments for all user activities except
the $i$'th activity for user $u$. \setcounter{equation}{7}
\begin{equation}
\label{equ:sampleFunction} \small \centering
\begin{aligned}
& P(z_{u,i}|\textbf{$z_{\neg u,i}$},~\textbf{w},~\textbf{v},~\textbf{s}, \circleddash) \\
\propto \quad & \alpha_{u,s_{u,i},l_{u,i},z_{u,i}} \times
\prod_{n=1}^{|{W}_{v_{u,i}}|}\beta_{z_{u,i},w_{u,i,n}}  \times
\gamma_{z_{u,i},v_{u,i}}
\end{aligned}
\end{equation}

In the M-step, we optimize the parameters $\circleddash$ to maximize
the log likelihood of the objective function with all  topic assignments
fixed. To update the parameters, we use the gradient descent
learning algorithm PSSG (Projected Scaled Sub-Gradient)~\cite{Schmidt:2007:LGM:1619797.1619850},
which is designed to solve optimization problems with L1
regularization on parameters. More importantly, PSSG is scalable
because it uses the quasi-Newton strategy with line search that is
robust to common functions. Let $L$ be the log-likelihood of the
model.  According to the limited-memory BFGS~\cite{Liu:1989:LMB:81100.83726} updates for the
quasi-Newton method, the gradients of model parameters  $\theta^0$,
$\theta^{user}$, $\theta^{native}$ and $\theta^{tourist}$ are
provided as follows. \begin{equation} \label{equ:updateTheta0}
\small \frac{\partial L}{\partial \theta
^{0}_{z}}=d(z)-\sum_{u=1}^{|U|}\sum_{i=1}^{|D_u|}\alpha_{u,s_{u,i},l_{u,i},z}
\end{equation}
\begin{equation}
\label{equ:updateThetaUser} \small \frac{\partial L}{\partial \theta
^{user}_{u,z}}=d(u,z)-\sum_{i=1}^{|D_u|}\alpha_{u,s_{u,i},l_{u,i},z}
\end{equation}
\begin{equation}
\label{equ:updateThetaNative} \small \frac{\partial L}{\partial
\theta
^{native}_{l,z}}=(1-s)\times(d(l,z)-\sum_{j=1}^{|D(l)|}\alpha_{u_j,s_j,l,z})
\end{equation}
\begin{equation}
\label{equ:updateThetaTourist} \small \frac{\partial L}{\partial
\theta
^{tourist}_{l,z}}=s\times(d(l,z)-\sum_{j=1}^{|D(l)|}\alpha_{u_j,s_j,l,z})
\end{equation}
  where $d(z)$ is the number of activities assigned to topic $z$, $d(u,z)$ represents how many activities are assigned to topic $z$ in $D_u$, $d(l,z)$ denotes the number of activities assigned to topic $z$ at location $l$, and  $D(l)$ is the set of activities occurring at the location $l$.

Similarly, the gradients of model parameters $\phi^0$,
$\phi^{topic}$, $\psi^0$, and $\psi^{topic}$  are computed as
follows:
\begin{equation}
\label{equ:updatePhi0} \small \frac{\partial L}{\partial \phi
^{0}_{w}}=d(w)-\sum_{z=1}^{K}d(z)\times\beta_{z,w}
\end{equation}
\begin{equation}
\label{equ:updatePhiTopic} \small \frac{\partial L}{\partial \phi
^{topic}_{z,w}}=d(z,w)-d(z)\times\beta_{z,w}
\end{equation}
\begin{equation}
\label{equ:updatePsi0} \small \frac{\partial L}{\partial \psi
^{0}_{v}}=d(v)-\sum_{z=1}^{K}d(z)\times\gamma_{z,v}
\end{equation}
\begin{equation}
\label{equ:updatePsiTopic} \small \frac{\partial L}{\partial \psi
^{topic}_{z,v}}=d(z,v)-d(z)\times\gamma_{z,v}
\end{equation}
 where $d(w)$ is the number of activities where the word $w$ appears,  and $d(z,w)$ is the number of activities where the word $w$ is assigned to the topic $z$.
 $d(v)$ is the number of activities associated with item $v$, and $d(z,v)$ represents the number of activities in which topic $z$ is assigned to item $v$.

 \subsection{Spatial Smoothing}

\begin{figure}[!htb]
      \centering
       \small
      \includegraphics[width=0.75\columnwidth]{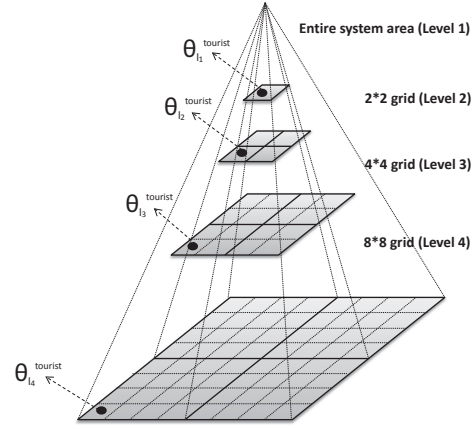}
     \vspace{-2mm}
  \caption{The Spatial Pyramid}
     \vspace{-3mm}
  \label{fig:pyramid}
  \end{figure}

To combat data sparsity when modelling the \emph{native preference} and \emph{tourist preference}, we adopt a tree structure, called spatial pyramid, to partition and index the entire geographic area. The spatial pyramid is  proposed in \cite{levandoski:lars}, which is constructed by partitioning item locations into spatial regions of varying sizes at different hierarchies. More specifically, the spatial pyramid decomposes the space into $H$ levels. The level $0$ has only one grid cell. For a given level $h$, the space is partitioned into $4^h$  grid cells of equal area. Thus, the space can be divided recursively into numerous cells at different levels with different granularity.  The graphical representation of the spatial pyramid is illustrated in Figure \ref{fig:pyramid}.

One of the fundamental assumptions in spatial data mining is that everything is related to everything else but nearby things are more related than distant things, which is proposed in \cite{gale:philosophy} as the first law of geography \cite{shekhar:data}. This law is known as the ``spatial autocorrelation''. The spatial pyramid structure can encode this law in an effective manner. That is,  for each location $l$, it can be represented by a path from the root node to its corresponding leaf node. We use a vector to describe the path, that is, ($l_{1}$, $l_{2}$,\dots, $l_{h}$, \dots, $l_{H}$). Based on the vector representation of a location, we can easily compute the proximity between two locations.  For example, if two locations in the spatial pyramid share more ancestors, then these two locations are more proximate.

When user activity data at a location $l$ is very sparse,
both the native preference $\theta^{native}_{l}$ and the tourist
preference $\theta^{tourist}_{l}$ may not be estimated accurately. To address this issue, we
exploit geographical correlation to enhance the prior knowledge about the model parameters  $\theta^{native}_l$ and $\theta^{tourist}_{l}$.  Intuitively, if two locations $l$ and $l'$ are proximate in the geographical space, then their local preferences $\theta^{native}_{l}$ and  $\theta^{native}_{l'}$ should be similar to each other.  The intuition can be applied to the tourist preferences $\theta^{tourist}_{l}$ and $\theta^{tourist}_{l'}$ similarly.  To integrate the information of geographical correlation into our Geo-SAGE model, we apply the \emph{additive} framework \cite{ahmed:latent,kanagal:supercharging} to compute native preference $\theta^{native}_l$ and tourist preference $\theta^{tourist}_l$ at location $l$ based on the path vector representation. Specifically, given a location $l$, its native preference and tourist preference are represented as follows.
 \begin{equation}
\small
\label{equ:localPreference} \centering
 \theta^{native}_{l}= \sum_{h=1}^{H}\theta^{native}_{l_{h}}
\end{equation}
 \begin{equation}
\small
\label{equ:localPreference6} \centering
 \theta^{tourist}_{l}= \sum_{h=1}^{H}\theta^{tourist}_{l_{h}}
\end{equation}
 According to the above equations, both the native preference and the tourist preference of a location depend on all of its ancestors up to the root. This representation method enables neighboring locations to share a similar preference as desired, i.e., the preferences are smoothed over the spatial pyramid. Meanwhile, if there are few or no activities at a location, we can still infer its preference guided by its ancestors.  Besides, once \emph{native preference} and \emph{tourist preference} for each level are learned, this modeling makes the switch over among various granularity fast and convenient by changing the lowest level in the model without re-training the parameters.

In the M-step of the model inference, we optimize the \emph{native
preference} and \emph{tourist preference} using PSSG. Because the two parameters are computed as the sum of the corresponding parameters at each level,
we need to calculate the gradients of the parameters for each level. Thus, we extend
Equations~\ref{equ:updateThetaNative} and
\ref{equ:updateThetaTourist}, which infer \emph{native preference}
and \emph{tourist preference} respectively, as follows.
\begin{equation}
\label{equ:updateThetaNativeSpatial} \small \frac{\partial
L}{\partial \theta
^{native}_{l_h,z}}=(1-s)\times(d(l_h,z)-\sum_{j=1}^{|D(l_h)|}\alpha_{u_j,s,l_j,z})
\end{equation}
\begin{equation}
\label{equ:updateThetaTouristSpatial} \small \frac{\partial
L}{\partial \theta ^{tourist}_{l_h,z}}=s \times
(d(l_h,z)-\sum_{j=1}^{|D(l_h)|}\alpha_{u_j,s,l_j,z})
\end{equation}
where $d(l_h,z)$ denotes the number of activity records at the
location $l_h$ assigned to topic $z$,  $D(l_h)$ is the set of
activity records occurring at the location $l_h$, and $u_j$ denotes
the user who generates the $j$-th activity record.

\subsection{Spatial Item Recommendation}

Our proposed Geo-SAGE model is employed to recommend spatial items
as follows. Given a user $u$ and his/her target location $l$, our
task is to recommend top-$k$ spatial items  that
$u$ wish to visit from the items
that the user has not visited before. More precisely, given the user and his/her target location, we first compute the indicator $s$ according to the distance between $u$'s home location and $l$.  After that, for each spatial item $v$  located at $l$ which has not been visited by $u$, we compute
  the probability that user $u$ visits  $v$ as follows.
 \begin{equation}
\label{equ:recommendation} \centering
\small
\begin{aligned}
& P(v, {W}_v|u,s,\circleddash) = \quad \sum_{z=1}^{K} P(v,{W}_v,z|u,s,\circleddash)\\
= & \sum_{z=1}^{K}P(z|u,s,\theta^0, \theta^{user}, \theta^{native}, \theta^{tourist}) \\
& \times P({W}_v|z, \phi^0, \phi^{topic})
 \times P(v|z, \psi^0, \psi^{topic})\\
= & \sum_{z=1}^{K}\alpha_{u,s,l,z} \times
\bigg{(}\prod_{n=1}^{|{W}_v|}\beta_{z,w_{v,n}}\bigg{)}^{\frac{1}{|W_v|}}
\times \gamma_{z,v}
\end{aligned}
\end{equation}
where $W_v$ denotes the content words describing item $v$. We adopt
the geometric mean for the probability of topic $z$ generating the word set
$W_v$, considering that the number of words associated with
different spatial items may be different.

By building a tree structure for the space, our model supports users to switch between different scales of the geo-region conveniently (e.g., zoom in/out on Google Map \cite{bao:location}), by switching between different levels in the tree structure. Although previous work also supports the change to granularity \cite{Yin:2013:LLR:2487575.2487608}, it needs to re-train all model parameters for the new granularity while our model does not.
\vspace{-8pt}
\section{Experiments}
In this section, we first describe the settings of experiments and
then demonstrate the  experimental results.

\subsection{Experimental Settings}
\subsubsection{Data Sets}

We perform experiments on two real large-scale LBSNs datasets,
Foursquare and Twitter.

\begin{figure*}[!htb]
\centering
 \small
\subfigure[Foursquare]{
\includegraphics[width=7cm,height=4cm]{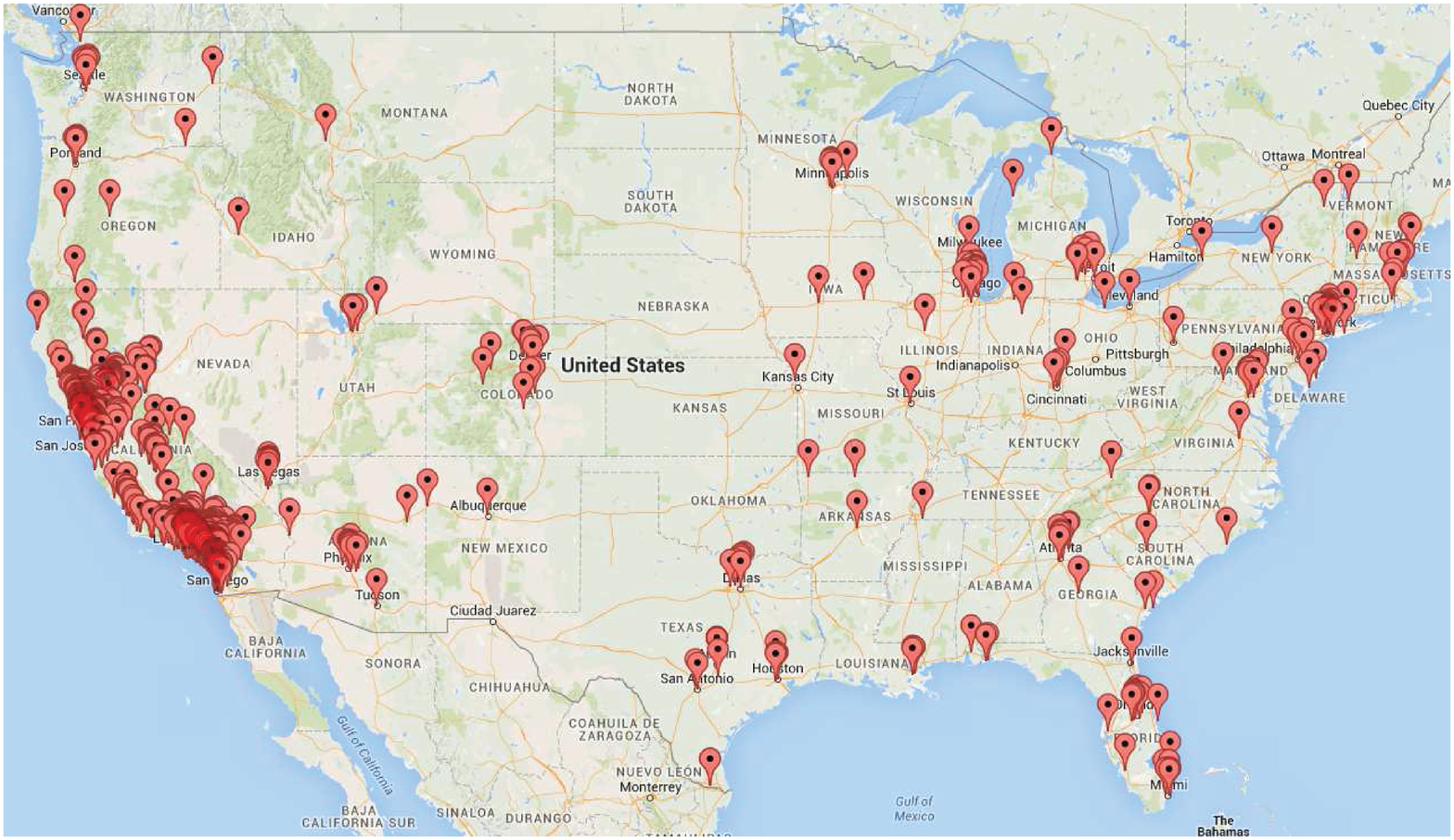}
\label{fig:foursquare} }
\subfigure[Twitter]{
 \includegraphics[width=7cm,height=4cm]{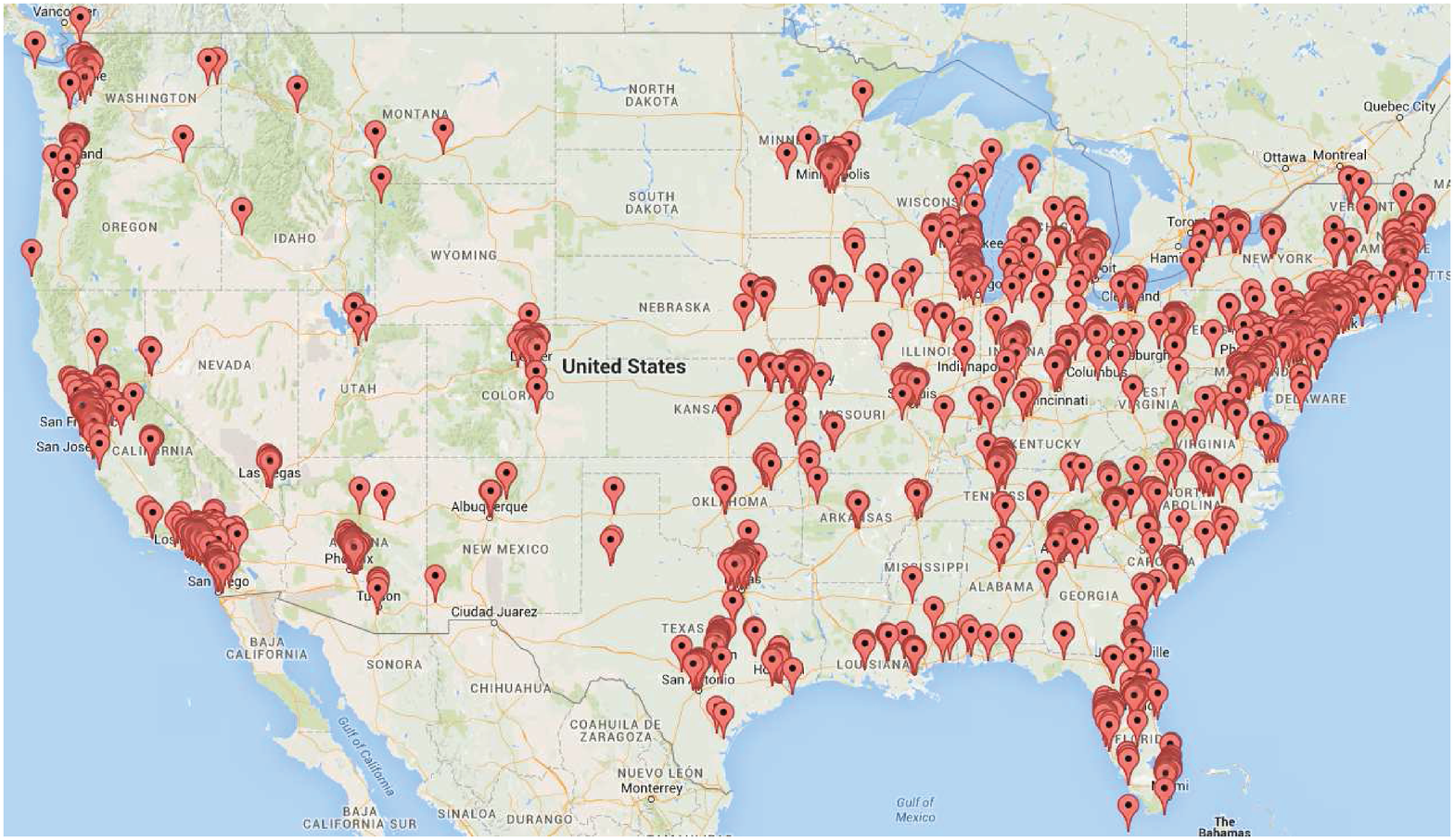}
 \label{fig:twitter}
}
\vspace{-3mm}
\caption{Distribution of User Check-in
Activities} \label{fig:maps}
\vspace{-5mm}
\end{figure*}

\textbf{Foursquare}.  This dataset contains the check-in history of
4,163 users who live in California, USA. For each user, it
contains the social networks, home location,  check-in venue identifiers,  location of
each venue in terms of latitude and longitude, and the content of
each check-in venue. The total number of check-ins in this dataset
is 483,813. Each check-in is stored as \emph{user-ID, venue-ID,
venue-location, venue-content, user-role}. Each record in social networks is stored as
\emph{userID, friendID} and the total number of social relationship
is 32,512. This dataset is publicly
available~\footnote{http://www.public.asu.edu/~hgao16/dataset.html}.
The distribution of the check-in activities is described in Figure
\ref{fig:foursquare}. From the distribution, we can see that, for the
users who live in California, although most of the check-ins occur in California, there is still a number of check-ins
occurring in other states. This is an evidence of the significance
of out-of-town recommendation.

\textbf{Twitter}.  This dataset is based on the publicly available
twitter dataset~\cite{conf/icwsm/ChengCLS11}. Twitter supports third-party location sharing services like Foursquare and Gowalla, where users of these services opt-in to share their check-ins on Twitter. But the original dataset does not contain the category or tag information about venues. So, we crawled the category and tag information associated with each venue from Foursquare with the help of its publicly available API \footnote{https://developer.foursquare.com/}.  The enhanced dataset contains 114,058 users and 1434,668 check-in activities. Each check-in record has the same format with the above Foursquare dataset. Nonetheless, this dataset does not contain user social network information.  Figure \ref{fig:twitter} describes the distribution of the check-in activities across the USA.  We observe that 18.22\% of check-in activities occur in California, and there are less than 7\% of activity records occurring in each of the other states. As the home locations of users in this dataset are not explicitly given, we adopt the method in~\cite{conf/icwsm/ScellatoNLM11} by inferring the user's home location as the cell in the spatial pyramid with the most check-ins.

To make the experiments repeatable, we make the Twitter dataset and our code publicly available~\footnote{http://net.pku.edu.cn/daim/yinhongzhi/index.html}.

\subsubsection{Comparative Approaches}

 We compare our Geo-SAGE model with the following three methods representing the state-of-the-art spatial item recommendation techniques.

 \textbf{LCA-LDA}. LCA-LDA is a location-content-aware recommender model which is developed to support spatial item recommendation for users traveling in new cities~\cite{Yin:2013:LLR:2487575.2487608}. This model takes into account both personal interests and local preferences of each city by exploiting both item co-visiting patterns and content of spatial items.  Compared with Geo-SAGE, LCA-LDA is a traditional mixture model which introduces ``switching'' variables to consider multiple factors. Besides, LCA-LDA ignores the roles of users and  does not distinguish  \emph{tourist preference} from \emph{native preference}.

   \textbf{UPS-CF}.  UPS-CF, proposed in~\cite{Ference:2013:LRO:2505515.2505637}, is a collaborative recommendation framework  which is especially designed for out-of-town users. This framework integrates user-based collaborative filtering and social-based collaborative filtering. That is, it recommends POIs to a target user according to the activity records of both his/her friends and similar users.

  \textbf{CKNN}.  CKNN~\cite{bao:location} projects a user's activity history into the category space and models user preference using a weighted category hierarchy. When receiving a query, CKNN retrieves all users and items located in the querying location, formulates a user-item matrix online, and then applies a user-based CF method to predict the rating of a querying user on an unvisited item. Note that the similarity between two users in CKNN is computed according to their weights in the category hierarchy, making CKNN a hybrid recommendation method.

  To further validate the benefits brought by exploiting the native preference or the tourist preference, distinguishing between tourist preference and native preference,  and  spatial smoothing based on the spatial pyramid, respectively, we implement three variant versions of our Geo-SAGE model.

   \textbf{Geo-SAGE-S1} is the simplified version of Geo-SAGE model which does not consider the native preference or tourist preference. Thus, this model cannot adapt to user interest drift. In this model,  $\alpha$ in Equation~\ref{eq:notations} is reformulated  as:
     \begin{equation}
     \small
     \alpha_{u,z} = \frac{
     exp(\theta^0_{z}+\theta^{user}_{u,z})}{\sum_{zz}
     exp(\theta^0_{zz}+\theta^{user}_{u,zz})}\nonumber
     \end{equation}

 \textbf{Geo-SAGE-S2} is the simplified version of Geo-SAGE where we ignore the roles of users and
       do not distinguish between tourist preference and native preference, similar to LCA-LDA~\cite{Yin:2013:LLR:2487575.2487608}. In this model,  $\alpha$ in Equation~\ref{eq:notations} is reformulated  as:

       \begin{equation}
       \small
        \alpha_{u,l,z} = \frac{
        exp(\theta^0_{z}+\theta^{user}_{u,z}+\theta^{crowd}_{l,z})}{\sum_{zz}
        exp(\theta^0_{zz}+\theta^{user}_{u,zz}+\theta^{crowd}_{l,zz})}\nonumber
       \end{equation}

   \textbf{Geo-SAGE-S3} is the simplified version of Geo-SAGE which does not exploit the geographical correlation in the spatial pyramid. Thus, the inferred native preference and tourist preference for location $l$ are not reliable when there are few or even no user activity records.

\subsubsection{Evaluation Methods}

Since our Geo-SAGE model is designed for both home-town recommendation and out-of-town recommendation, we evaluate the recommendation effectiveness of our model under the two scenarios respectively.
Given a user profile in terms of a collection of user activities, we divide the user's activities into a training set and a test set. For the scenario of home-town recommendation, we randomly select 30\% of the activity records occurring at the user's home town as test set, and use the remaining activity records as the training set. Similarly, for the scenario of out-of-town recommendation, we randomly select 30\% of the activity records generated by the user when he/she travels out of town as the test set, and use the remaining activity records as training set. To decide whether an activity record occurs when the user is in his/her home town or not, we measure the location distance between the user's home town and the spatial item (e.g., $|$$l_u$ $-$ $l_v$$|$). If the distance is greater than $100km$, then we assume the activity occurs when the user is out-of-town. The threshold $d=100km$ is selected because a distance around $100km$ is the typical human radius of ``reach'', which takes about 1 to 2 hours to drive.

According to the above dividing strategies, we split the
user activity dataset $D$ into the training  set $D_{train}$ and
the test set $D_{test}$. To evaluate the recommendation methods, we
adopt the evaluation methodology and measurement Recall@$k$, which is applied
in~\cite{Hu:2013:STM:2507157.2507174,yin:challenging,Yin:2013:LLR:2487575.2487608}. Specifically, for each user activity  record $(u,$ $~v,$ $~l_v,$ $~W_v$, $s)$  in $D_{test}$:\\ 1) we compute the ranking
score for spatial item $v$ and  all other spatial items which
are within the circle of radius $d$ centered at $l_v$ and unvisited
by $u$ previously; 2) we form a ranked list by ordering all of these
spatial items according to their ranking scores. Let $r$ denote the
position of the spatial item $v$ within this list. The best result
corresponds to the case where  $v$ precedes all the unvisited
spatial items (that is, $r=1$); and 3) we form a top-$k$ recommendation
list by picking the $k$ top ranked spatial items from the list. If
$r\leq k$, we have a hit (i.e., the ground truth item  $v$ is
recommended to the user). Otherwise, we have a miss.

 The computation of Recall@$k$ proceeds as follows. We define hit@$k$ for
a single test case as either the value 1, if the test item $v$
appears in the top-$k$ results, or the value $0$, if otherwise. The overall
Recall@$k$ is defined by averaging over all test cases:
\vspace{-3pt}
\begin{equation}\label{eq:18}
\vspace{-3pt} \small
    Recall@k=\frac{\#hit@k}{|D_{test}|}\nonumber
    \vspace{-3pt}
\end{equation}

where $\#hit@k$ denotes the number of hits in the test set, and
$|D_{test}|$ is the number of all test cases.

\subsection{Recommendation Effectiveness}
\begin{figure}[!htb]
\centering
 \small
\subfigure[Out-of-Town]{
 \includegraphics[width=4cm,height=3.5cm]{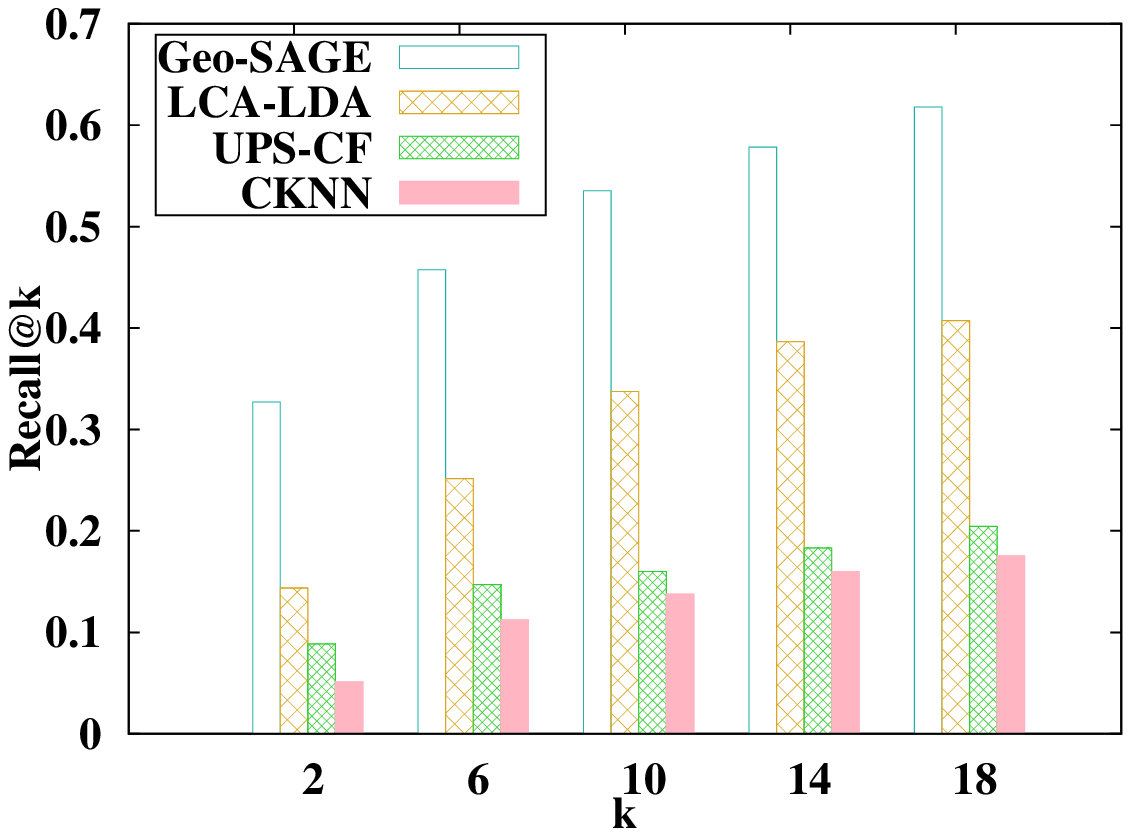}
 \label{fig:foutoftown}
}
\subfigure[Home-Town]{
\includegraphics[width=4cm,height=3.5cm]{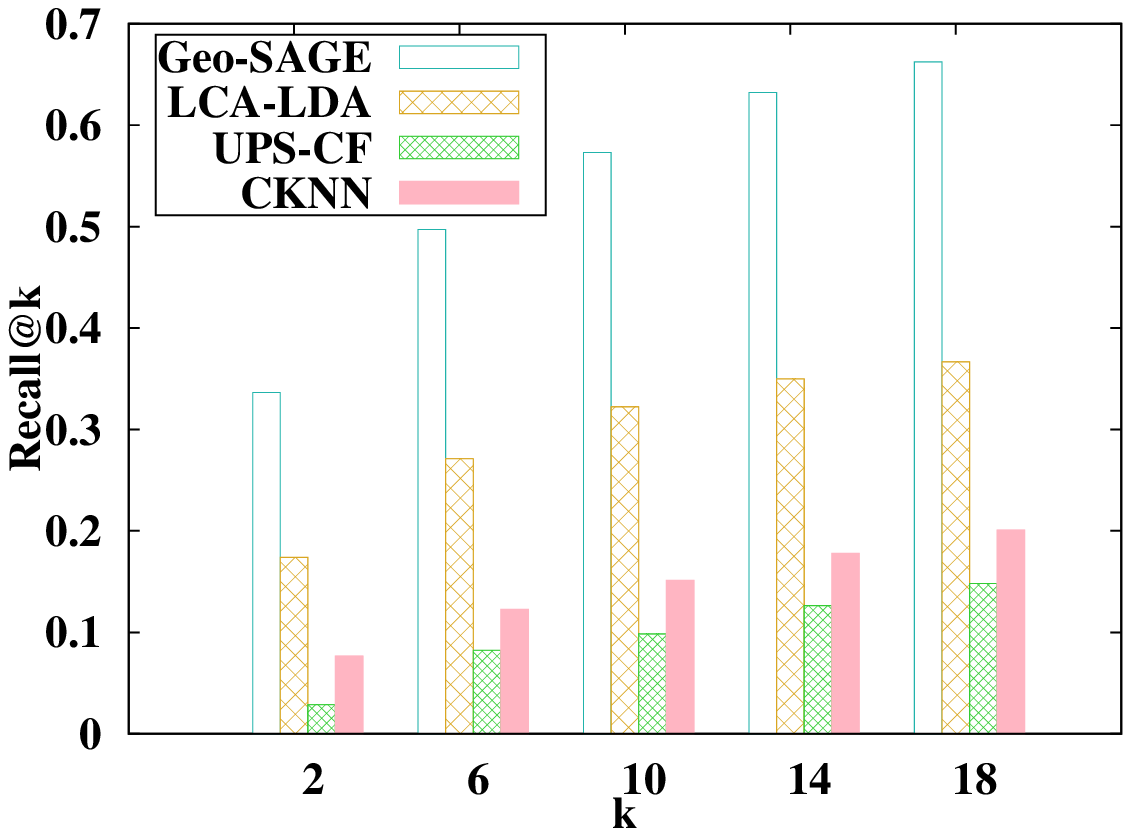}
\label{fig:fintown}
}
\vspace{-3mm}
\caption{Performance on Foursquare Dataset}
\label{fig:resultFoursquare}
\vspace{-3mm}
\end{figure}

\begin{figure}[!htb]
\centering
 \small
\subfigure[Out-of-Town]{
 \includegraphics[width=4cm,height=3.5cm]{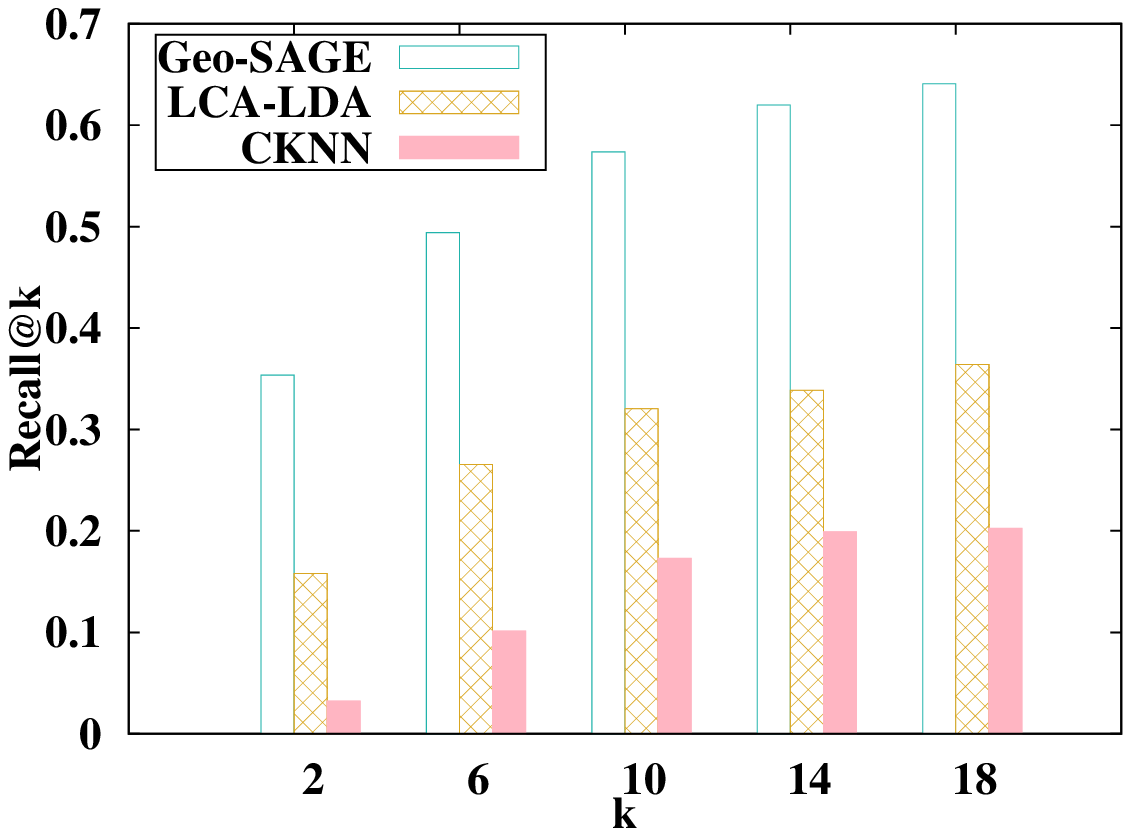}
 \label{fig:toutoftown}
}
\subfigure[Home-Town]{
\includegraphics[width=4cm,height=3.5cm]{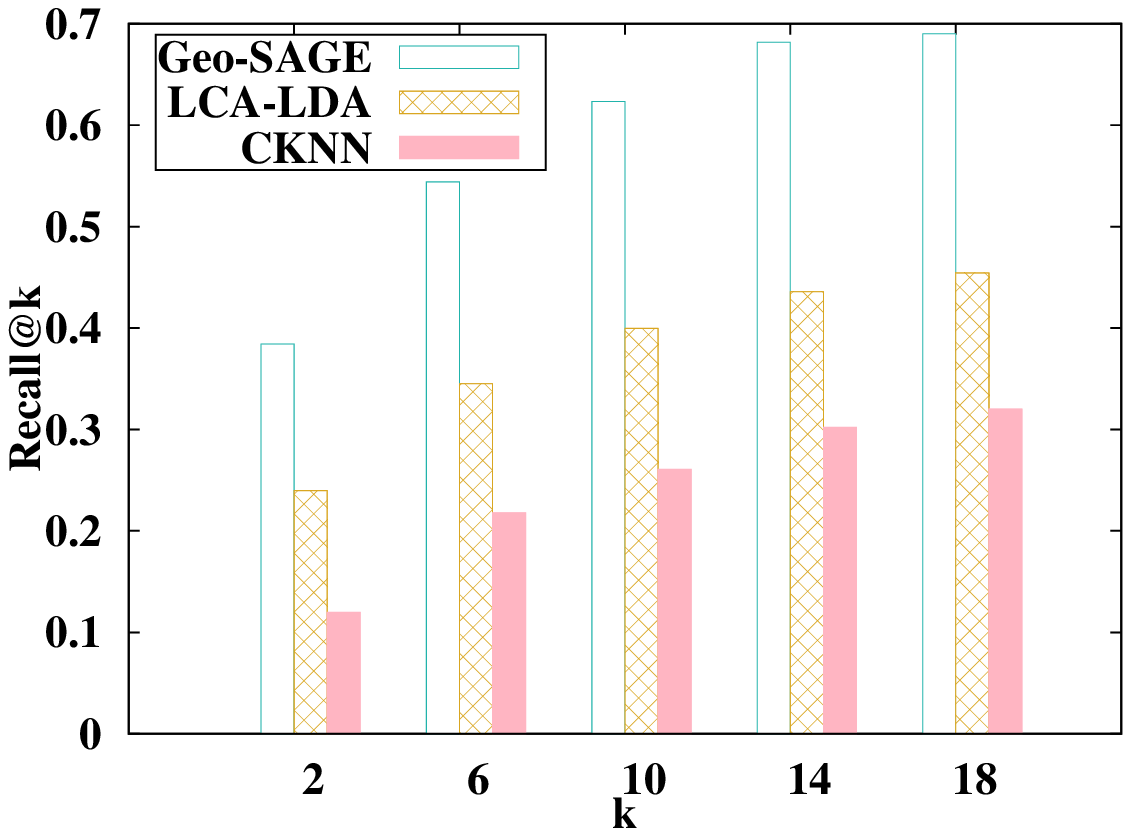}
\label{fig:tintown}
}
\vspace{-3mm}
\caption{Performance on Twitter Dataset}
\label{fig:resultTwitter}
\vspace{-5mm}
\end{figure}

In this part, we  present experimental results of the comparing recommendation methods with well-tuned parameters. Figures~\ref{fig:resultFoursquare} and~\ref{fig:resultTwitter}  report the recommendation effectiveness  on the Foursquare and Twitter datasets, respectively. From the figures, we observe that the recall values gradually rise with respect to the increase of $k$. This is because, by returning more spatial items, it is more likely to discover the ones that users would like to visit. Note that we show only the performance when $k$ is set to $2,6,10,14,18$. A greater value of
$k$ is usually ignored for the top-$k$ recommendation task.

It is apparent that the recommendation methods have significant performance disparity in terms of the top-$k$ recall.  Figure~\ref{fig:foutoftown} presents the recommendation accuracy  in the scenario of out-of-town recommendation where the recall of Geo-SAGE is about 0.327 when $k=2$, and 0.535 when $k=10$. This means that Geo-SAGE has a probability of 32.7\% to place an appealing point of interest  in the top-2 recommendations and 53.5\% to place it in the top-10 recommendations. Clearly, our proposed Geo-SAGE model outperforms other competitor models significantly, demonstrating the advantages of Geo-SAGE over other competitor methods. Several observations are made from the results: 1) Geo-SAGE and LCA-LDA  perform much better than CKNN. This is because both of them exploit the crowd's  behaviors at the target region, which effectively alleviates the problem of user interest drift, while CKNN does not consider the issue since the local experts discovered by CNKK have same or similar interests with target users. 2)  UPS-CF drops behind Geo-SAGE and LCA-LDA, showing the advantages of using latent topic models to capture  users' interests by exploiting the content of their visited spatial items. Through the medium of content, Geo-SAGE and LCA-LDA  transfer the users' interests inferred at home town to out-of-town regions. In contrast, UPS-CF is a mixture of collaborative filtering and social filtering, which ignores the effect of content. Besides, according to the recent survey in~\cite{cho:friendship}, for movement farther than 100km from home location, the probability of visiting the exact same location as a friend has visited in the past is low. 3) Geo-SAGE achieves much higher recall value than LCA-LDA, showing the advantages of sparse additive modeling over the traditional mixture modeling. Additionally, the superiority of Geo-SAGE is partly due to the distinguishing between the native preference and the tourist preference.

Figure \ref{fig:resultTwitter} shows the recommendation effectiveness on the Twitter dataset. As the social relationship is unavailable on this dataset, UPS-CF has not been evaluated on this dataset. The trend of the comparison result is similar to that presented in Figure \ref{fig:resultFoursquare}. In both settings, Geo-SAGE performs the best, followed by LCA-LDA.

There are two parameters in Geo-SAGE, namely, the height of spatial pyramid ($H$)  and the number of topics ($K$).  The experimental results
presented above are obtained with the optimal parameter settings:
(1) the optimal height of spatial pyramid is $5$ for both Foursquare and Twitter datasets; (2) the optimal values of $K$ are 50 for the Foursquare dataset, and 100 for the Twitter dataset.

\subsection{Impact of Different Factors}

\begin{figure}[!t]
\centering
 \small
\subfigure[Out-of-Town]{
 \includegraphics[width=4cm,height=3.5cm]{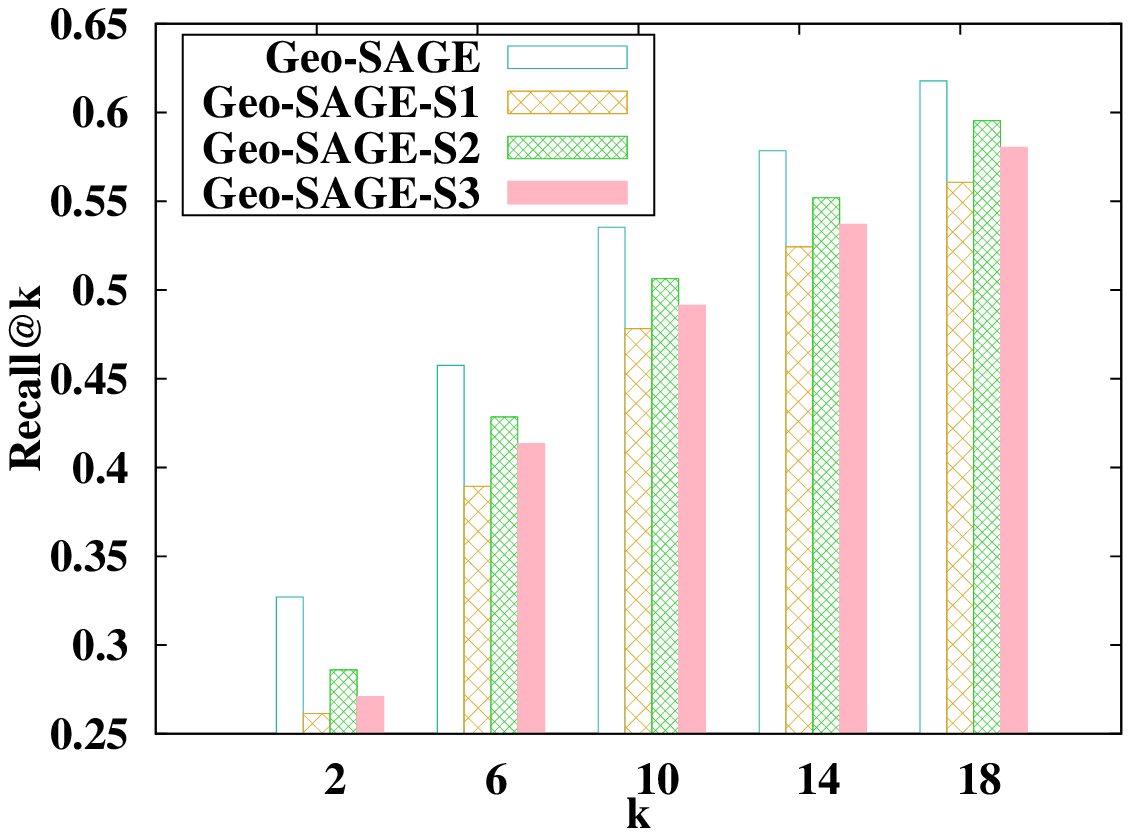}
 \label{fig:foutoftown1}
}
\subfigure[Home-Town]{
\includegraphics[width=4cm,height=3.5cm]{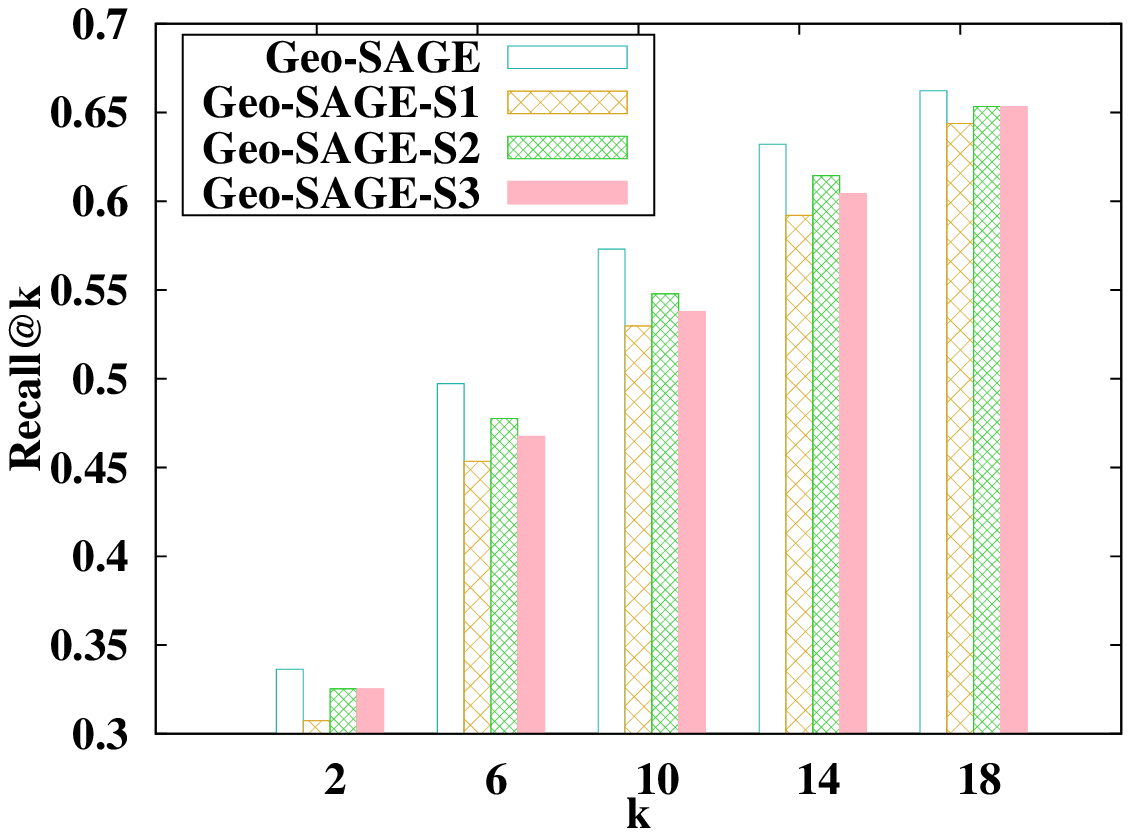}
\label{fig:fintown1}
}
\vspace{-3mm}
\caption{Impact of Different Factors on Foursquare Dataset}
\label{fig:resultFoursquare1}
\vspace{-3mm}
\end{figure}

\begin{figure}[!t]
\centering
 \small
\subfigure[Out-of-Town]{
 \includegraphics[width=4cm,height=3.5cm]{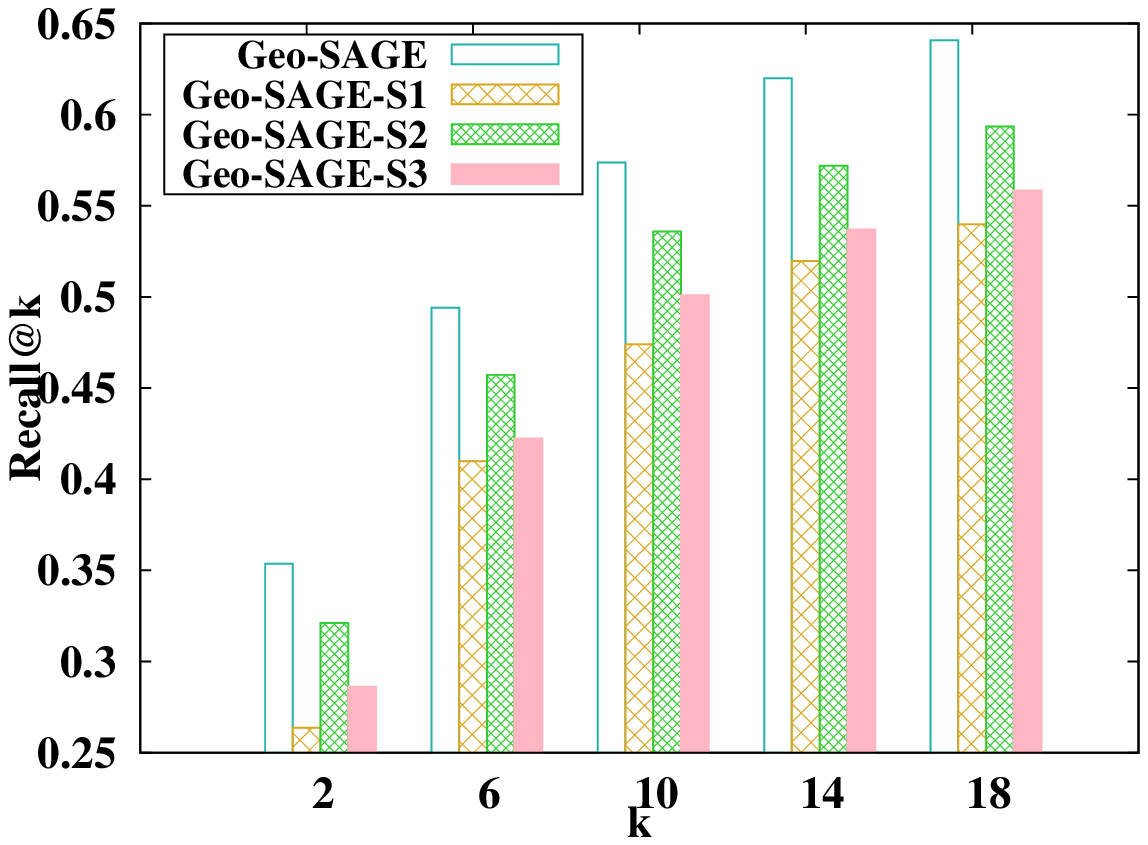}
 \label{fig:toutoftown1}
}
\subfigure[Home-Town]{
\includegraphics[width=4cm,height=3.5cm]{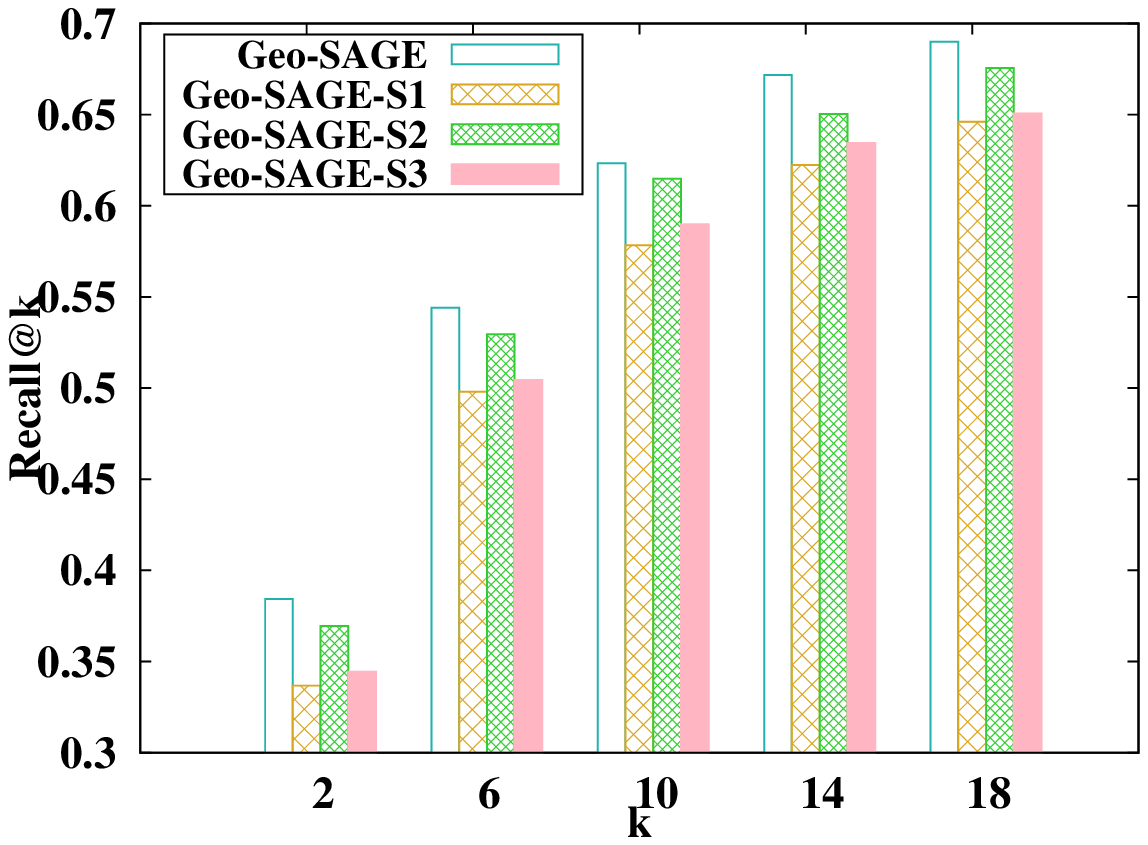}
\label{fig:tintown1}
}
\vspace{-3mm}
\caption{Impact of Different Factors on Twitter Dataset}
\label{fig:resultTwitter1}
\vspace{-5mm}
\end{figure}

To  validate the benefits brought by exploiting the native preference or the tourist preference, distinguishing between tourist preference and native preference, as well as spatial smoothing based on the spatial pyramid, respectively, we compare Geo-SAGE with the three variant versions: Geo-SAGE-S1, Geo-SAGE-S2 and Geo-SAGE-S3. The comparison results are shown in Figures~\ref{fig:resultFoursquare1} and \ref{fig:resultTwitter1}.  From the results, we observe that Geo-SAGE consistently outperforms the three variant versions for both out-of-town recommendation  and home-town recommendation, indicating the benefits brought by each factor, respectively. For instance, the performance gap between Geo-SAGE and Geo-SAGE-S3 validates the effectiveness of exploiting the geographical correlation and smoothing both native and tourist preferences over the spatial pyramid to alleviate data sparsity.
 We also observe that Geo-SAGE-S2 and Geo-SAGE-S3 outperform Geo-SAGE-S1 consistently, showing the advantages brought by exploiting the crowd's  activity histories when individual user's behavior data is sparse.   Another observation is that the performance gap between our Geo-SAGE model and the three variants in the task of home-town recommendation is smaller than that in the task of out-of-town recommendation, showing that the performance differences among these four methods become less obvious when people travel in home town where  the issue of data sparsity is not serious.

\begin{table}[!htb]
{
\small
\centering
\begin{tabular}{|c|c|c|c|c|c|c|}
  \hline
  \multirow{2}{*}{Height} & \multicolumn{3}{c|}{Foursquare}& \multicolumn{3}{c|}{Twitter}
  \\\cline{2-7}& Re@2&Re@10&Re@20& Re@2&Re@10&Re@20\\\hline
    2 &  0.282&0.511&0.616&0.287&0.459&0.602\\\hline
   3 &  0.301&0.519&0.620&0.292&0.537&0.615 \\\hline
   4 & 0.319&0.528&0.629&0.318&0.559&0.635\\\hline
   \textbf{5} & \textbf{0.327}&\textbf{0.535}&\textbf{0.637}&\textbf{0.353}&\textbf{0.573}&\textbf{0.653}\\\hline
   6 &  0.296&0.504&0.605&0.289&0.555&0.618\\\hline
   7 &  0.270&0.465&0.567&0.256&0.537&0.582\\\hline
\end{tabular}
\vspace{-3mm}
\caption{Impact of Height of Spatial Pyramid}
\label{tab:h}
\vspace{-3mm}
}
\end{table}

 To further study the effect of spatial pyramid in Geo-SAGE, we tried different parameter setups for the height of the spatial pyramid.  Tuning this parameter is critical to the performance of this model. We tested the performance of Geo-SAGE model by varying the height of spatial pyramid from $1$ to $7$, and present the results in Table~\ref{tab:h}. Due to space constraint, we only show the experimental results for the out-of-town recommendation. From the results, we observe that as the height of the spatial pyramid increases, the $Recall$ values of Geo-SAGE first increase, and then decrease.
  One possible reason for early increasing $Recall$ values is that increasing the height increases the exploitation of spatial effect and makes the inference of native and tourist preferences more precise.  Later on, $Recall$ decreases as the height of spatial pyramid gets larger, because increasing the height makes users' activity data in a region cell sparser.  Geo-SAGE achieves its best performance  when the height of spatial pyramid is set to 5 on the two datasets, which could be a tradeoff of aforementioned two factors.

\vspace{-3pt}
\subsection{Test for Cold Start Problem}

\begin{figure}[!t]
	\centering
\small
		\subfigure[With Less than 5 Activities]{
			\includegraphics[width=4cm,height=3.5cm]{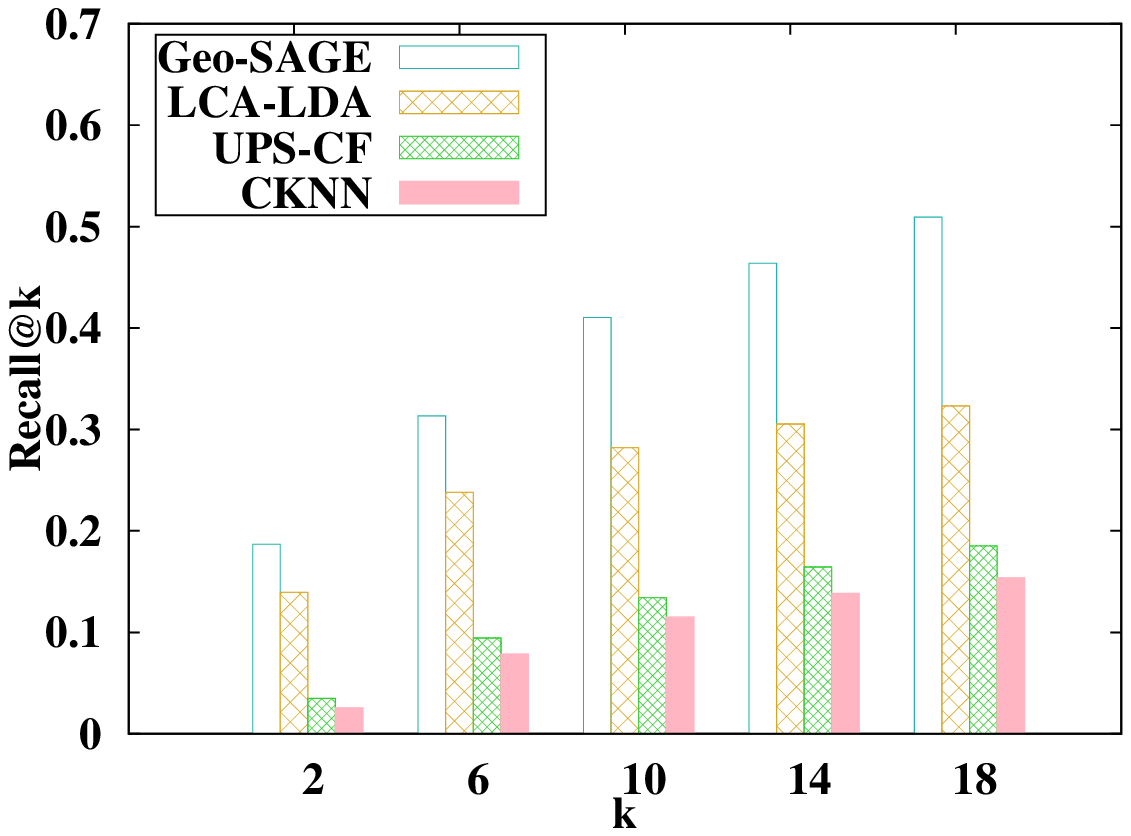}
			\label{fig:coldstart5}
		}
	\subfigure[With 0 Activities]{
		\includegraphics[width=4cm,height=3.5cm]{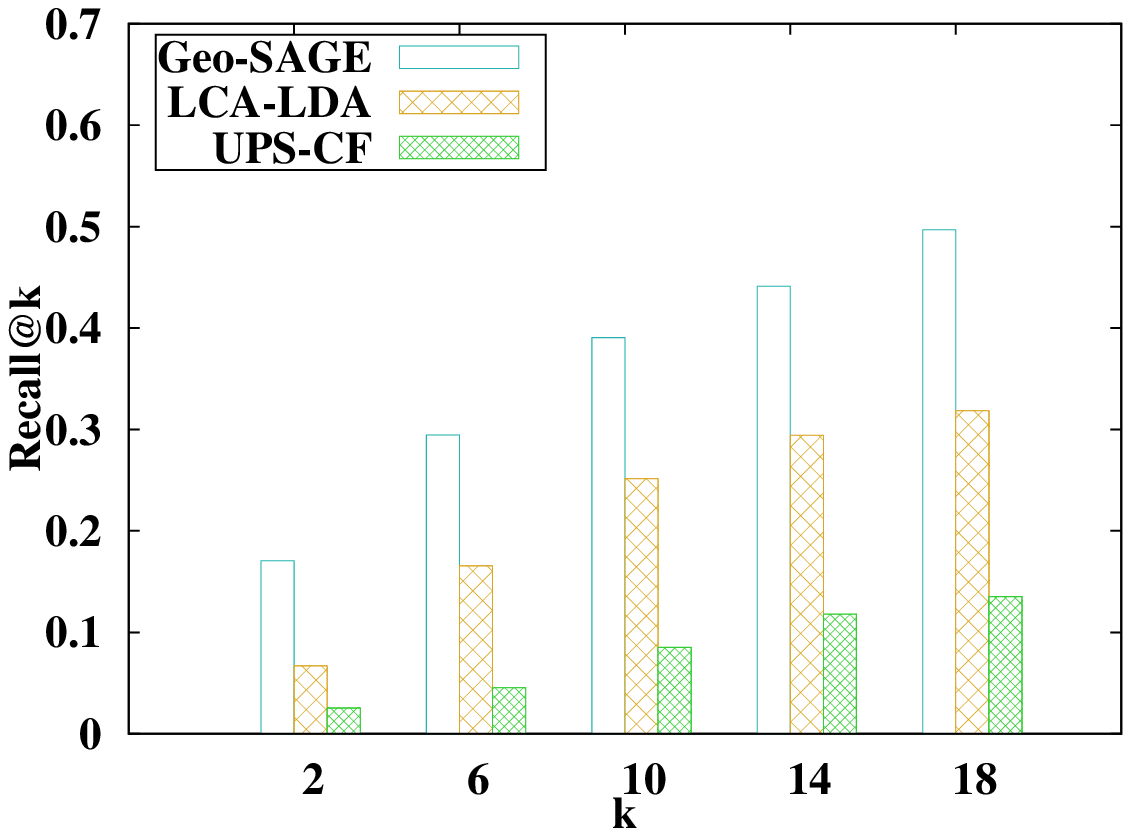}
		\label{fig:coldstart0}
	}
	\vspace{-4mm}
		\caption{Recommendation for Cold-start Users}
	\label{fig:coldstart}
	\vspace{-3mm}
\end{figure}

We further conduct experiments to study the effectiveness of different recommendation algorithms handling the cold start problem. Cold start is a critical problem in the domain of recommendation, where users have not visited any spatial item, or have visited only a few items. We test the recommendation effectiveness for cold-start users on the Foursquare dataset and present the results in Figure~\ref{fig:coldstart}. Figure~\ref{fig:coldstart}(a) reports the result for users with less than 5 activity records,  and Figure~\ref{fig:coldstart}(b) reports the result for users without any activity record.

 By comparing Figures~\ref{fig:resultFoursquare} and \ref{fig:coldstart}, we observe that, although the recommendation effectiveness of all algorithms decrease to various degrees, our proposed Geo-SAGE still performs the best. Note that it is very hard to capture cold-start users' preference from extremely limited number of spatial items visited by target users. That is why CKNN performs worst by using user activity history only.  Geo-SAGE, LCA-LDA and UPS-CF overcome the scarce of user activity history by exploiting the wisdom of the crowd and social friends, respectively. The behaviors of the crowd or social friends at the target region supply many valuable references and clues that are potentially useful for  spatial item recommendations.  Geo-SAGE outperforms LCA-LDA due to the fact that Geo-SAGE distinguishes the tourist preference from the native preference, and a user is more likely to visit the spatial items preferred by the crowd who have the same role as him/her at the target region. Note that all users' home town information is known in the Foursquare dataset. Thus, for users with no activity records, Geo-SAGE can also identify whether a user is a tourist or a local and then recommend the most popular items preferred by the tourists or the locals in the target region.  Another observation is that both Geo-SAGE and LCA-LDA outperform UPS-CF  significantly. This is because UPS-CF  is a memory-based method which suffers from severe data sparsity, while Geo-SAGE and LCA-LDA  are latent class models that can handle the data sparsity problem to a great extent by reducing dimension and integrating content information.

\begin{figure}[!t]
   \centering
   \small
   \includegraphics[width=0.6\columnwidth]{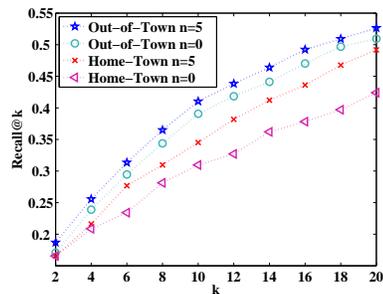}
   \centering
 \vspace{-5mm}
\caption{Home-town and Out-of Town Recommendation for Cold-start Users}
\label{fig:fcscompare}
\vspace{-6mm}
\end{figure}

Figure \ref{fig:fcscompare} illustrates the performance of Geo-SAGE for cold-start users in home-town and out-of-town recommendation. $n$ denotes the maximum number of activity records that each cold-start user has in the training set.  From the results, we observe that for cold-start users,  Geo-SAGE achieves better performance in the task of out-of-town recommendation than in the task of home-town recommendation.  Besides, as $n$ increases from $0$ to $5$, Geo-SAGE achieves a larger performance increase in home-town recommendation than in out-of-town recommendation. The results indicates that for cold-start users, the tourist preference plays a more important role in improving out-of-town recommendation than the native preference does in improving home-town recommendation. This is because users' visiting behaviors at home town are more personalized and diverse,  while their out-of-town behaviors tend to be more consistent. That is, different users tend to visit the same spatial items when they travel in the same out-of-town region. On the other hand, the results indicate that users' personal interests have more influence on home-town recommendation than on out-of-town recommendation. That is why home-town recommendation is more sensitive to the number of user historical activities.

\vspace{-8pt}
\section{Related Work}

Spatial item recommendation, also called location or place recommendation, has been considered as an essential task in the domain of recommender system. Recently, with the easy access
of large-scale user activity records in LBSNs, many recent work has tried to improve spatial  recommendation by exploiting and integrating geographical and social influence, temporal effect and content information of POIs.

\textbf{Geo-Social Influence}.  Geo-Social influence indicates  that people tend to explore nearby POIs of a POI that they or their friends have visited before~\cite{Ye:2011:EGI:2009916.2009962}. Many recent studies~\cite{Lian:2014:GJG:2623330.2623638,DBLP:conf/aaai/ChengYKL12,Ye:2011:EGI:2009916.2009962,cho:friendship} showed that there is a strong correlation between user check-in activities and  geographical distance as well as social connections, so most of current POI recommendation work mainly focus on leveraging the geographical and social influences to improve recommendation accuracy.  For example, Ye et al.~\cite{Ye:2011:EGI:2009916.2009962} delved into POI recommendation by investigating the geographical influences among locations and proposed a system that combines user preferences,  social influence and geographical influence.  Cheng et al.~\cite{DBLP:conf/aaai/ChengYKL12} investigated the geographical influence through combining a multi-center Gaussian model, matrix factorization and social influence together for location recommendation.  Lian et al.~\cite{Lian:2014:GJG:2623330.2623638} incorporated spatial clustering phenomenon resulted by geographical influence into a weighted matrix factorization framework to deal with the matrix sparsity. Ference et al.~\cite{Ference:2013:LRO:2505515.2505637} designed a collaborative recommendation framework which considers the activity records generated by both friends and similar users in a mixture way.

\textbf{Temporal Effect}.  The temporal effect of user check-in activities in LBSNs has also attracted much attention from researchers.  POI recommendation with temporal effect mainly leverages temporal cyclic patterns and temporal sequential patterns on LBSNs.  Gao et al.~\cite{Gao:2013:ETE:2507157.2507182} investigated the temporal cyclic patterns of user check-ins in terms of temporal non-uniformness and temporal consecutiveness. Yuan et al.~\cite{Yuan:2013:TPR:2484028.2484030} also incorporated the temporal cyclic information into a user-based collaborative filtering framework for time-aware POI recommendation.  Cheng et al.~\cite{Cheng:2013:YLG:2540128.2540504} introduced the task of successive personalized POI recommendation in LBSNs by embedding the temporal sequential patterns.

\textbf{Content Information}. Most recently, researchers explored the content information of POIs to alleviate the problem of data sparsity.  Hu et al.~\cite{Hu:2013:STM:2507157.2507174} proposed a spatial topic model for POI recommendation considering both spatial aspect and textual aspect of  user posts from Twitter.    Yin et al.~\cite{Yin:2013:LLR:2487575.2487608} exploited both personal interests and local preferences based on the contents associated with spatial items.  Liu et al.~\cite{Liu:2013:SDM} studied the effect of POI-associated tags for POI recommendation with an aggregated LDA and matrix factorization method.  Gao et al.~\cite{HUIJI2015AAAI}
  studied both POI-associated contents and user sentiment information (e.g., user comments) into POI recommendation and reported their good performance.

As described above, while there are many studies to alleviate the data sparsity by exploiting geographical-social influence, temporal effect and content information, they did not address the challenges arising from either \emph{travel locality} or \emph{interest drift}  for the out-of-town recommendation.  For example, most of the above work assumed that users are in their home towns, they did not consider user interest drift across regions. Our work in this paper distinguishes from previous work in several points. First, to the best of our knowledge, we are the first to  simultaneously address the three challenges in a unified model. Second, although both~\cite{Yin:2013:LLR:2487575.2487608} and ~\cite{liu:learning} exploited the local crowd's preferences, they ignored the user's role and did not distinguish native preference from tourist preference. Third, we proposed a novel effective method to represent and infer both native and tourist preferences based on a well-designed spatial index structure.

\vspace{-8pt}
\section{Conclusion and Future Work}

In this paper, we proposed a geographical sparse
additive generative model, Geo-SAGE,  for spatial item recommendation, which effectively overcomes the challenges arising from data sparsity,
 travel locality and interest drift.  Specifically, to combat data sparsity and travel locality, Geo-SAGE exploited both the co-occurrence pattern of spatial items and their content to infer and transfer user interests.  To address interest drift,  Geo-SAGE incorporated the
 native or tourist  preference at the target location.  To alleviate the  data sparsity confronted by the inference of native preference and tourist preference for each region, Geo-SAGE employed an additive framework to smooth the preferences over a well-designed spatial pyramid.    We conducted extensive experiments to evaluate the performance of our Geo-SAGE model on two real large-scale datasets. The experimental results revealed the advantages of the Geo-SAGE model over other spatial item recommendation methods, for both home-town and out-of-town recommendations, demonstrating the effectiveness of Geo-SAGE in facilitating users to travel in their home towns as well as in regions they are not familiar with.

 As a promising research direction,  we would like to explore enhancements to our  model by integrating the social influence and temporal effect.

\bibliographystyle{abbrv}
\small
\bibliography{reference}
\end{document}